\pdfoutput=1 
\documentclass[useAMS,usenatbib]{mnras}       

\usepackage{graphicx}
\usepackage{times}

\usepackage{eso-pic}

\usepackage{hyperref}
\usepackage{amssymb}
\usepackage{amsmath}
\usepackage{lineno}
\usepackage{setspace}

\def\aap{{A\&A}}

\def\apjl{{ApJL}}

\def\apjs{{ApJS}}

\newcommand{\beq}{\begin{equation}}
\newcommand{\eeq}{\end{equation}}
\newcommand{\beqa}{\begin{eqnarray}}
\newcommand{\eeqa}{\end{eqnarray}}

\newcommand{\mocks}{{\rm {DES-BAO-MOCKS}}}
\newcommand{\main}{{\rm {DES-BAO-MAIN}}}
\newcommand{\methodAPS}{{\rm {DES-BAO-$\ell$-METHOD}}}
\newcommand{\methodACF}{{\rm {DES-BAO-$\theta$-METHOD}}}
\newcommand{\methodXi}{{\rm {DES-BAO-$s_{\perp}$-METHOD}}}
\newcommand{\photoz}{{\rm {DES-BAO-PHOTOZ}}}

\usepackage{xspace}
\newcommand{\change}[1]{{\textcolor{black}{#1}}\xspace}

\begin{document}


\title[Galaxy sample for DES Y1 BAO measurements]{Dark Energy Survey Year 1 Results:
   Galaxy Sample for BAO Measurement}
\author[Crocce et al.]{
\parbox{\textwidth}{
\Large
M.~Crocce$^{1}$\thanks{e-mail: martincrocce@gmail.com},
A.~J.~Ross$^{2}$,
I.~Sevilla-Noarbe$^{3}$,
E.~Gaztanaga$^{1}$,
J.~Elvin-Poole$^{4}$,
S.~Avila$^{5,6}$,
A.~Alarcon$^{1}$,
K. ~C.~Chan$^{1,48}$,
N.~Banik$^{7}$,
J.~Carretero$^{8}$,
E.~Sanchez$^{3}$,
W.~G.~Hartley$^{9,10}$,
C.~S{\'a}nchez$^{8}$,
T.~Giannantonio$^{11,12,13}$,
R.~Rosenfeld$^{14,15}$,
A.~I.~Salvador$^{3}$,
M.~Garcia-Fernandez$^{3}$,
J.~Garc\'ia-Bellido$^{7}$,
T.~M.~C.~Abbott$^{16}$,
F.~B.~Abdalla$^{17,10}$,
S.~Allam$^{7}$,
J.~Annis$^{7}$,
K.~Bechtol$^{18}$,
A.~Benoit-L{\'e}vy$^{19,10,20}$,
G.~M.~Bernstein$^{21}$,
R.~A.~Bernstein$^{22}$,
E.~Bertin$^{20,19}$,
D.~Brooks$^{10}$,
E.~Buckley-Geer$^{7}$,
A.~Carnero~Rosell$^{23,15}$,
M.~Carrasco~Kind$^{24,25}$,
F.~J.~Castander$^{1}$,
R.~Cawthon$^{26}$,
C.~E.~Cunha$^{27}$,
C.~B.~D'Andrea$^{22}$,
L.~N.~da Costa$^{23,15}$,
C.~Davis$^{27}$,
J.~De~Vicente$^{3}$,
S.~Desai$^{28}$,
H.~T.~Diehl$^{7}$,
P.~Doel$^{10}$,
A.~Drlica-Wagner$^{7}$,
T.~F.~Eifler$^{29,30}$,
P.~Fosalba$^{1}$,
J.~Frieman$^{7,26}$,
J.~Garc\'ia-Bellido$^{6}$,
D.~W.~Gerdes$^{31,32}$,
D.~Gruen$^{27,33}$,
R.~A.~Gruendl$^{25,24}$,
J.~Gschwend$^{23,15}$,
G.~Gutierrez$^{7}$,
D.~Hollowood$^{34}$,
K.~Honscheid$^{2,35}$,
B.~Jain$^{22}$,
D.~J.~James$^{36}$,
E.~Krause$^{30,29}$,
K.~Kuehn$^{37}$,
S.~Kuhlmann$^{38}$,
N.~Kuropatkin$^{7}$,
O.~Lahav$^{10}$,
M.~Lima$^{39,15}$,
M.~A.~G.~Maia$^{15,23}$,
J.~L.~Marshall$^{40}$,
P.~Martini$^{2,41}$,
F.~Menanteau$^{24,25}$,
C.~J.~Miller$^{31,32}$,
R.~Miquel$^{42,8}$,
R.~C.~Nichol$^{5}$,
W.~J.~Percival$^{5}$,
A.~A.~Plazas$^{29}$,
M.~Sako$^{22}$,
V.~Scarpine$^{7}$,
R.~Schindler$^{33}$,
D.~Scolnic$^{26}$,
E.~Sheldon$^{43}$,
M.~Smith$^{44}$,
R.~C.~Smith$^{16}$,
M.~Soares-Santos$^{7,45}$,
F.~Sobreira$^{15,46}$,
E.~Suchyta$^{47}$,
M.~E.~C.~Swanson$^{25}$,
G.~Tarle$^{32}$,
D.~Thomas$^{5}$,
D.~L.~Tucker$^{7}$,
V.~Vikram$^{38}$,
A.~R.~Walker$^{16}$,
B.~Yanny$^{7}$,
Y.~Zhang$^{7}$
\begin{center} (The Dark Energy Survey Collaboration) \end{center}
}
\vspace{0.4cm}
\\
\parbox{\textwidth}{
\scriptsize
$^{1}$ Institute of Space Sciences (ICE, CSIC) \& Institut d'Estudis Espacials de Catalunya (IEEC), Campus UAB, Carrer de Can Magrans, s/n,  08193 Barcelona, Spain\\
$^{2}$ Center for Cosmology and Astro-Particle Physics, The Ohio State University, Columbus, OH 43210, USA\\
$^{3}$ Centro de Investigaciones Energ\'eticas, Medioambientales y Tecnol\'ogicas (CIEMAT), Madrid, Spain\\
$^{4}$ Jodrell Bank Center for Astrophysics, School of Physics and Astronomy, University of Manchester, Oxford Road, Manchester, M13 9PL, UK\\
$^{5}$ Institute of Cosmology \& Gravitation, University of Portsmouth, Portsmouth, PO1 3FX, UK\\
$^{6}$ Instituto de Fisica Teorica UAM/CSIC, Universidad Autonoma de Madrid, 28049 Madrid, Spain\\
$^{7}$ Fermi National Accelerator Laboratory, P. O. Box 500, Batavia, IL 60510, USA\\
$^{8}$ Institut de F\'{\i}sica d'Altes Energies (IFAE), The Barcelona Institute of Science and Technology, Campus UAB, 08193 Bellaterra (Barcelona) Spain\\
$^{9}$ Department of Physics, ETH Zurich, Wolfgang-Pauli-Strasse 16, CH-8093 Zurich, Switzerland\\
$^{10}$ Department of Physics \& Astronomy, University College London, Gower Street, London, WC1E 6BT, UK\\
$^{11}$ Institute of Astronomy, University of Cambridge, Madingley Road, Cambridge CB3 0HA, UK\\
$^{12}$ Kavli Institute for Cosmology, University of Cambridge, Madingley Road, Cambridge CB3 0HA, UK\\
$^{13}$ Universit\"ats-Sternwarte, Fakult\"at f\"ur Physik, Ludwig-Maximilians Universit\"at M\"unchen, Scheinerstr. 1, 81679 M\"unchen, Germany\\
$^{14}$ ICTP South American Institute for Fundamental Research \& Instituto de F\'{\i}sica Te\'orica, Universidade Estadual Paulista, S\~ao Paulo, Brazil\\
$^{15}$ Laborat\'orio Interinstitucional de e-Astronomia - LIneA, Rua Gal. Jos\'e Cristino 77, Rio de Janeiro, RJ - 20921-400, Brazil\\
$^{16}$ Cerro Tololo Inter-American Observatory, National Optical Astronomy Observatory, Casilla 603, La Serena, Chile\\
$^{17}$ Department of Physics and Electronics, Rhodes University, PO Box 94, Grahamstown, 6140, South Africa\\
$^{18}$ LSST, 933 North Cherry Avenue, Tucson, AZ 85721, USA\\
$^{19}$ CNRS, UMR 7095, Institut d'Astrophysique de Paris, F-75014, Paris, France\\
$^{20}$ Sorbonne Universit\'es, UPMC Univ Paris 06, UMR 7095, Institut d'Astrophysique de Paris, F-75014, Paris, France\\
$^{21}$ Department of Physics and Astronomy, University of Pennsylvania, Philadelphia, PA 19104, USA\\
$^{22}$ Observatories of the Carnegie Institution of Washington, 813 Santa Barbara St., Pasadena, CA 91101, USA\\
$^{23}$ Observat\'orio Nacional, Rua Gal. Jos\'e Cristino 77, Rio de Janeiro, RJ - 20921-400, Brazil\\
$^{24}$ Department of Astronomy, University of Illinois at Urbana-Champaign, 1002 W. Green Street, Urbana, IL 61801, USA\\
$^{25}$ National Center for Supercomputing Applications, 1205 West Clark St., Urbana, IL 61801, USA\\
$^{26}$ Kavli Institute for Cosmological Physics, University of Chicago, Chicago, IL 60637, USA\\
$^{27}$ Kavli Institute for Particle Astrophysics \& Cosmology, P. O. Box 2450, Stanford University, Stanford, CA 94305, USA\\
$^{28}$ Department of Physics, IIT Hyderabad, Kandi, Telangana 502285, India\\
$^{29}$ Jet Propulsion Laboratory, California Institute of Technology, 4800 Oak Grove Dr., Pasadena, CA 91109, USA\\
$^{30}$ Department of Astronomy/Steward Observatory, 933 North Cherry Avenue, Tucson, AZ 85721-0065, USA\\
$^{31}$ Department of Astronomy, University of Michigan, Ann Arbor, MI 48109, USA\\
$^{32}$ Department of Physics, University of Michigan, Ann Arbor, MI 48109, USA\\
$^{33}$ SLAC National Accelerator Laboratory, Menlo Park, CA 94025, USA\\
$^{34}$ Santa Cruz Institute for Particle Physics, Santa Cruz, CA 95064, USA\\
$^{35}$ Department of Physics, The Ohio State University, Columbus, OH 43210, USA\\
$^{36}$ Event Horizon Telescope, Harvard-Smithsonian Center for
Astrophysics, MS-42, 60 Garden Street, Cambridge, MA 02138, UK \\
$^{37}$ Australian Astronomical Observatory, North Ryde, NSW 2113, Australia\\
$^{38}$ Argonne National Laboratory, 9700 South Cass Avenue, Lemont, IL 60439, USA\\
$^{39}$ Departamento de F\'isica Matem\'atica, Instituto de F\'isica, Universidade de S\~ao Paulo, CP 66318, S\~ao Paulo, SP, 05314-970, Brazil\\
$^{40}$ George P. and Cynthia Woods Mitchell Institute for Fundamental Physics and Astronomy, and Department of Physics and Astronomy, Texas A\&M University, College Station, TX 77843,  USA\\
$^{41}$ Department of Astronomy, The Ohio State University, Columbus, OH 43210, USA\\
$^{42}$ Instituci\'o Catalana de Recerca i Estudis Avan\c{c}ats, E-08010 Barcelona, Spain\\
$^{43}$ Brookhaven National Laboratory, Bldg 510, Upton, NY 11973, USA\\
$^{44}$ School of Physics and Astronomy, University of Southampton,  Southampton, SO17 1BJ, UK\\
$^{45}$ Department of Physics, Brandeis University, Waltham, MA 02453, USA\\
$^{46}$ Instituto de F\'isica Gleb Wataghin, Universidade Estadual de Campinas, 13083-859, Campinas, SP, Brazil\\
$^{47}$ Computer Science and Mathematics Division, Oak Ridge National Laboratory, Oak Ridge, TN 37831\\
$^{48}$ School of Physics and Astronomy, Sun Yat-Sen University, Guangzhou 510275, China \\
}
}

\date{Prepared for submission to MNRAS}
\pagerange{\pageref{firstpage}--\pageref{lastpage}} \pubyear{2017}
\label{firstpage}
\maketitle

\begin{abstract}
We define and characterise a sample of 1.3 million galaxies extracted from the first
year of Dark Energy Survey data, optimised to measure Baryon Acoustic
Oscillations in the presence of significant redshift uncertainties. The
sample is dominated by luminous red galaxies located at redshifts $z
\gtrsim 0.6$. We define the exact selection using color and magnitude cuts that balance
the need of high number densities and small photometric redshift
uncertainties, using the corresponding forecasted BAO distance error
as a figure-of-merit in the process. The typical photo-$z$ uncertainty varies from
$2.3\%$ to $3.6\%$ (in units of 1+$z$) from $z=0.6$ to $1$, with
number densities from $200$ to $130$ galaxies per deg$^2$ in
tomographic bins of width $\Delta z = 0.1$. Next we summarise the
validation of the photometric redshift estimation. We characterise and
mitigate observational systematics including stellar contamination,
and show that the clustering on large scales is robust in front of those contaminants. 
We show that the clustering signal in the auto-correlations and
cross-correlations is generally consistent with theoretical models, which serves as an additional test of the redshift distributions.
\end{abstract}


\section{Introduction}
\label{sec:intro}

The use of the imprint of Baryon Acoustic Oscillations (BAO) in the spatial 
distribution of galaxies as a standard ruler has become one of the common 
methods in current observational cosmology to understand the Universe. The physics that causes BAO is 
well understood. Primordial perturbations generated acoustic waves in the 
photon-baryon fluid until decoupling $(z \sim 1100)$. These sound waves lead 
to the large oscillations observed in the power spectrum of the CMB 
anisotropies, but they are also visible in the clustering of matter, and 
therefore galaxies, as a high density region around the original source of 
the perturbation, at a distance given by the sound horizon length at
recombination. This high 
density region shows as a small excess in the number of pairs of galaxies separated 
by $\sim 150$ Mpc. Since the sound horizon is very precisely measured in the cosmic 
microwave background \citep{2016A&A...594A..13P}, the BAO measurements can be 
used as a standard ruler. This is, therefore, a geometrical probe of the expansion 
rate of the Universe, that maps the angular diameter distance and the Hubble 
parameter as functions of the redshift. There have now been multiple 
detections of the BAO in redshift surveys \citep{2005ApJ...633..560E,2010MNRAS.401.2148P,2015MNRAS.449..835R,2017MNRAS.470.2617A,2017arXiv170506373A,2015A&A...574A..59D,2017A&A...603A..12B,2001MNRAS.327.1297P,2005MNRAS.362..505C,2011MNRAS.415.2892B,2011MNRAS.416.3017B} and it is considered as one of the main cosmological probes for current and planned cosmological projects.

A key feature of the BAO method is the fact that the sound horizon length is large, 
and, therefore, very deep and wide galaxy surveys are needed in order to reach precise 
measurements of the BAO scale. But, at the same time, this large 
scale protects the BAO feature from large corrections due to astrophysical and 
non-linear effects of structure formation and therefore from systematic errors, making BAO a
solid probe of the expansion rate of the Universe.

The Dark Energy Survey (DES) is one of the most important of the currently ongoing 
large galaxy surveys and, as its name suggests, it is specially designed to attack 
the problem of the physical nature of the dark energy. It will do it using several 
independent and complementary methods at the same time. One of them is the precise 
study of the spatial distribution of galaxies, and in particular, the BAO standard 
ruler. DES is a photometric survey, which means that its precision in the measurement 
of redshifts is limited, preventing the measurement of the Hubble parameter 
evolution. However, the evolution of the angular distance with redshift is possible, 
through the measurement of angular correlation functions \citep{2003ApJ...598..720S,Pad05,2005MNRAS.363.1329B,Pad07,2011MNRAS.417.2577C,2011MNRAS.411..277S,2012MNRAS.419.1689C,2012ApJ...761...13S,2013MNRAS.435.3017D}. 

Although DES will only measure BAO in the angular distribution of galaxies, a 
determination of the photometric redshift as precise as possible
brings several benefits. It allows a finer tomography in the mapping of the BAO evolution
with the redshift and makes the analysis cleaner, reducing the correlations between
redshift bins. A sample of Luminous Red Galaxies (LRGs) would fit these 
requirements \citep{Pad05,Pad07}. LRGs are luminous and massive galaxies with a nearly uniform
Spectral Energy Distribution (SED), but with a strong 
break at 4000 \AA \ in the rest frame. These features allow a clean selection and an accurate determination of 
the redshift for this type of galaxies, even in photometric surveys. 
This selection has been done previously for imaging data at $z
\lesssim 0.6$ \citep{Pad05}. But the BAO
scale has already been measured with high precision in this redshift range
(e.g. \cite{2017MNRAS.470.2617A} and references therein).
In order to go to higher redshifts, the 
selection criteria need to be redefined. The 4000 \AA \, feature enters the i band at $z=0.75$, and
the methods used in previous selections are not valid anymore.

In this paper we describe the selection of a sample of red galaxies to measure BAO in
DES, that includes, but is not limited to, LRGs. The selection is defined by two
conditions. On the one hand, keep the determination of the photometric redshift
as precise as possible. On the other hand, keep the galaxy density high enough to
have a BAO measurement that is not limited by shot noise.

In order to guide our efforts to select an optimized sample for measuring BAO 
distance scales, we rely on Fisher matrix forecasts. \cite{SeoEis07} provide a 
framework and simple formulae to predict the precision that one can achieve with 
a given set of galaxy data.
Thus, we will test how Fisher matrix forecasts vary given the variations 
obtained for the number density and estimated redshift uncertainty given a set 
of color-magnitude cuts. 

This paper, detailing the BAO sample selection, is one of a series describing the supporting
work leading to the BAO measurement using DES Y1 data presented
in \cite{BAOmain} (hereafter \main). As part of such series, one paper
presents the mock galaxy catalogues, \cite{Halogen} (hereafter \mocks). 
\cite{Enrique} discusses in detail the photo-$z$ validation, and we
denote it \photoz. \cite{kwan}, from now on \methodACF, introduces the BAO extraction pipeline using a
tomographic analysis of angular correlation functions, while
\cite{camacho} presents the study of the angular power spectrum
(hereafter \methodAPS). Lastly,
\cite{2017MNRAS.472.4456R}, in what follows referred to as \methodXi, introduced a novel
technique to infer BAO distances using the three-dimensional
correlation function binned in projected separations.

This paper is organized as follow: in section ~\ref{sec:Y1data}, a description of the
main features of the DES-Y1 catalogue is given: in section~\ref{sec:sample_selection}, we
give a detailed description of the selection cuts that define the data sample that has
been used to measure the BAO scale in DES; section~\ref{sec:photozvalidation} contains a description
of the procedure that has been developed and applied in DES in order
to ensure the quality of the photometric redshift determination, and to determine its relation with
the true redshift; section \ref{sec:mask} describes the masking scheme and the treatment
of the variable depth in the survey; section \ref{sec:obs_syst} is a description
of the analysis and mitigation of observational systematic errors on the clustering measurement; and 
finally, section \ref{sec:two-point} describes the measured two-point correlation 
and cross-correlation functions and their evolution with redshift for the selected 
sample. We finish with our conclusions in section \ref{sec:conclusions}.

\begin{figure*}
\includegraphics[width=0.9\linewidth]{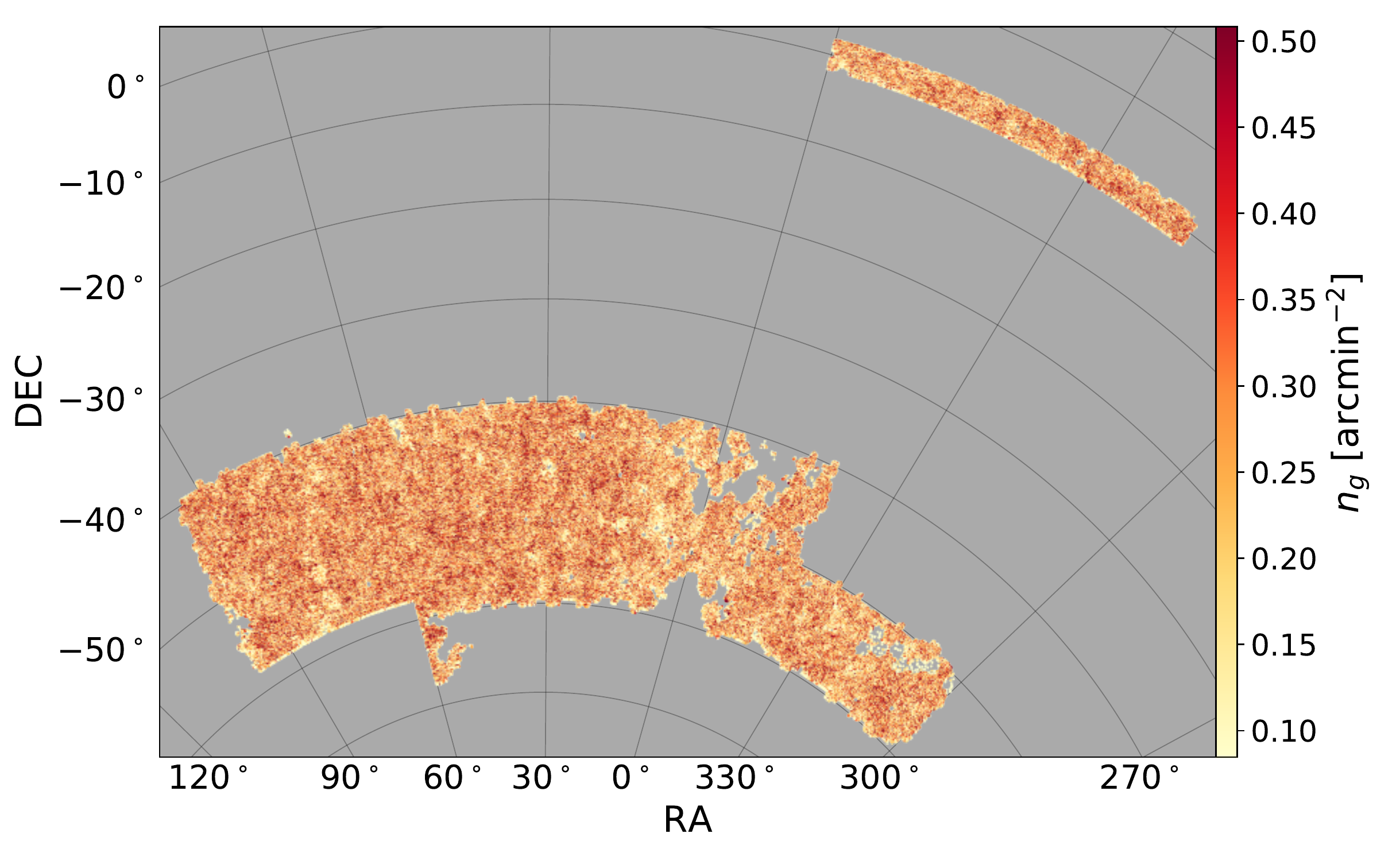}
\caption{Angular distribution and projected density of the DES-Y1 red galaxy sample described in this paper, and subsequently used for BAO measurements. The unmasked footprint comprises the two largest compact regions of the dataset: one in the southern hemisphere of 1203 deg$^2$, overlapping South Pole Telescope observations (SPT; Carlstrom et al. 2011), and 115 deg$^2$ near the celestial equator,  overlapping with Stripe 82 (S82, Annis et al. 2014). The sample consists of about $1.3$ million galaxies with photometric redshifts in the range $[ 0.6 - 1.0 ]$ and constitutes the baseline for our DES Y1 BAO analysis.} 
\label{Fig:footprint}
\end{figure*}

\begin{table*}
\centering
\caption{Complete description of the selection performed to obtain a sample dominated by red galaxies with a good compromise of photo-$z$ accuracy and number density, optimal for the BAO measurement presented in \main. The redshifts of the resulting catalogue are then computed using different codes (BPZ and DNF) as described in Sec \ref{sec:Y1data}. Therefore, any subsequent photo-$z$ selection can be done either with $z_{photo}$ from BPZ or DNF.}
\begin{tabular}{ccc}
\hline
\hline
Keyword & Cut & Description \\
\hline
Gold            & observations present in the Gold catalog & \cite{Y1Gold} \\
Quality         & {\tt flags\_badregion < 4; \tt flags\_gold = 0} & Sec.\ref{sec:mask}; Sec.\ref{sec:Y1data}  \\ 
Footprint       & 1336 deg$^2$ (1221 deg$^2$ in SPT and 115 deg$^2$ in S82) & Fig. \ref{Fig:footprint} Sec.\ref{sec:mask}  \\ 
Color\,Outliers & $ -1 < g_{\rm auto} - r_{\rm auto} < 3 $ & Sec. \ref{sec:fluxcuts}  \\ 
                & $ -1 < r_{\rm auto} - i_{\rm auto} < 2.5 $ & Sec. \ref{sec:fluxcuts} \\ 
                & $ -1 < i_{\rm auto} - z_{\rm auto} < 2 $ & Sec. \ref{sec:fluxcuts} \\ 
$[$Optimized$]$ Color Selection           & $(i_{\rm auto} - z_{\rm auto}) + 2.0(r_{\rm auto} - i_{\rm auto}) > 1.7$ & Sec. \ref{sec:cuts3-color} \\ 
$[$Optimized$]$ Completeness Cut    & $i_{\rm auto} < 22$ &  Sec. \ref{sec:fluxcuts} \\ 
$[$Optimized$]$ Flux Selection            & $17.5 <  i_{\rm auto} < 19.0 + 3.0z_{\rm BPZ-AUTO}$ & Sec. \ref{sec:cuts3-magnitude}  \\ 
Star-galaxy separation & {\tt spread\_model\_i} + (5/3) {\tt spreaderr\_model\_i} $> 0.007$ & Sec. \ref{sec:cuts2} \\ 
Photo-$z$ range & $[0.6 - 1.0$]& Sec. \ref{sec:photozvalidation}\\
\hline
\hline
\end{tabular}
\label{Tab:sampledef}
\end{table*}

\section{DES Y1 Data}
\label{sec:Y1data}

The BAO galaxy sample we will define in this work makes use of the
first year of data (Y1) from the Dark Energy Survey. This
photometric dataset has been produced using the Dark Energy Camera
(DECam, \cite{decam})
observations, processed and calibrated by the DES Data
Management system (DESDM) \citep{desdm1,desdm2,desdm3} and finally curated,
optimized and complemented into the Gold catalog (hereafter denoted `Y1GOLD'), as
described in \cite{Y1Gold}. For each band, single exposures are combined in \textit{coadds}
to achieve a higher depth. We keep track of
the complex geometry that the combinations of these dithered
exposures will create at each point in the sky in terms of
observing conditions and survey properties. Objects are
detected in chi-squared combinations of
the \textit{r}, \textit{i} and \textit{z} coadds to create
the final coadd catalog \citep{1999AJ....117...68S}.

Y1GOLD covers a total footprint of more than 1800 deg$^2$;
this footprint is defined by a {\sc Healpix} \citep{healpix}
map at resolution $N_{\mathrm{side}}=4096$ and includes only
area with a minimum total exposure time of at least 90
seconds in each of the \textit{griz} bands, and a valid calibration solution (see \cite{Y1Gold} for
details). This footprint is divided into several disjoint sub-regions which encompass the supernova survey areas, 
a region overlapping stripe 82 from the SDSS footprint (S82;
\cite{2014ApJ...794..120A}) 
and a larger area overlapping with the South Pole Telescope coverage
(SPT; \cite{SPTref}).
 Figure \ref{Fig:footprint} shows the angular distribution of galaxies, 
 selected as described in Section \ref{sec:sample_selection}, that includes these two areas. 
 A series of veto masks, including
masks for bright stars and the Large Magellanic Cloud among others,
reduce the area by $\sim 500$ deg$^2$, leaving $1336$ deg$^2$
suitable for LSS study. Other areas that are severely
affected by imaging artifacts or otherwise have a high
density of image artifacts are masked out as well. 
Section \ref{sec:mask} provides a full account of the final mask used
in combination with the final BAO sample. 
\textit{``Bad'' regions} information is propagated to the `object' level by using the \texttt{flags\_badregion} column in the
catalog. Finally, individual objects which have been
identified as being problematic by the DESDM processing or
by the vetting process carried out by the scientists in the
collaboration are flagged when configuring the catalog (this
is done through the \texttt{flags\_gold} column). All data we
describe in this and in subsequent sections are drawn from
quantities and maps released as part of the DES Y1 Gold
catalog and are fully described in \cite{Y1Gold}.

The photometry used in this work comes mainly from two
different sources: 
\begin{itemize}
\item the \texttt{SExtractor}
(\cite{sextractor}) \texttt{AUTO} magnitudes, which are derived from the best matched elliptical aperture according to the coadd object elongation and angle in the sky, measured using the coadded object flux;
\item Multi-Object Fitting (MOF) pipeline, which performs a multi-epoch and multi-band fit of the shape and per-band fluxes directly on the single epoch exposures for each of the coadd objects, with additional neighboring light subtraction. This is described in more detail in \cite{Y1Gold}.
\end{itemize}

Using these photometric measurements, we will consider three different photometric redshift catalogues. Two of them are built using BPZ  \citep{2000ApJ...536..571B}, a Bayesian template-fitting method, and another using  a machine learning approach: the Directional Neighborhood Fitting (DNF) algorithm as described in \cite{2016MNRAS.459.3078D}. They are combined with the photometric quantities described above and used as follows:

\begin{itemize}
{\item BPZ run with \texttt{AUTO} magnitudes (hereafter ${z_{\rm BPZ-AUTO}}$) used for making the selection of the overall sample.}
{\item BPZ run with MOF magnitudes (hereafter ${z_{\rm BPZ-MOF}}$)
  used for redshift binning and transverse distance calculation, finally used as secondary catalogue to
  show the robustness of the analysis.}
{\item DNF run with MOF magnitudes (hereafter ${z_{\rm DNF-MOF}}$) used for redshift binning and transverse distance calculation, finally used as our fiducial catalogue.}
\end{itemize}

We should note that BPZ with  \texttt{AUTO} magnitudes is part of the DESDM data
reduction pipeline and is available early on in the catalogue
making. This explains why we used that particular combination for
sample selection. We did not find, and do not expect, the relative optimization of the
sample selection and cuts to depend much on the particular photo-$z$
catalogue (but the final absolute error on BAO distance measurement does). 

In Section \ref{sec:photozvalidation}, we summarize the validation
performed to select and characterise the true redshift distributions of the binned samples,
which is described in detail in \photoz.   

Throughout our analysis we assume the redshift estimate of each galaxy to be the {\it mean} redshift of the redshift posterior for BPZ, or the predicted value for the object in the fitted hyper-plane from the DNF code (see \cite{2016MNRAS.459.3078D}. Any potential biases from these estimates are calibrated as described in Section \ref{sec:photozvalidation}.

\section{Sample Selection}
\label{sec:sample_selection}

In this section, we  describe the steps towards the construction of a
red galaxy dominated sample, optimized for BAO measurements, starting from the dataset described in Section \ref{sec:Y1data}. The selection is performed over the largest continuous regions of the survey at this point, namely SPT and S82. Objects are selected so that we avoid imaging artifacts and pernicious regions with foreground objects using the cuts on {\tt flags\_badregion} and {\tt flags\_gold} described therein. In the rest of this section we go into finer details on the flux, color and star-galaxy separation selection.  

In Table \ref{Tab:sampledef},  we summarise this sample selection, including references to the sections where these cuts are explained. 

\subsection{Completeness and color outliers cuts}
\label{sec:fluxcuts}

\begin{figure}
\includegraphics[width=0.9\linewidth]{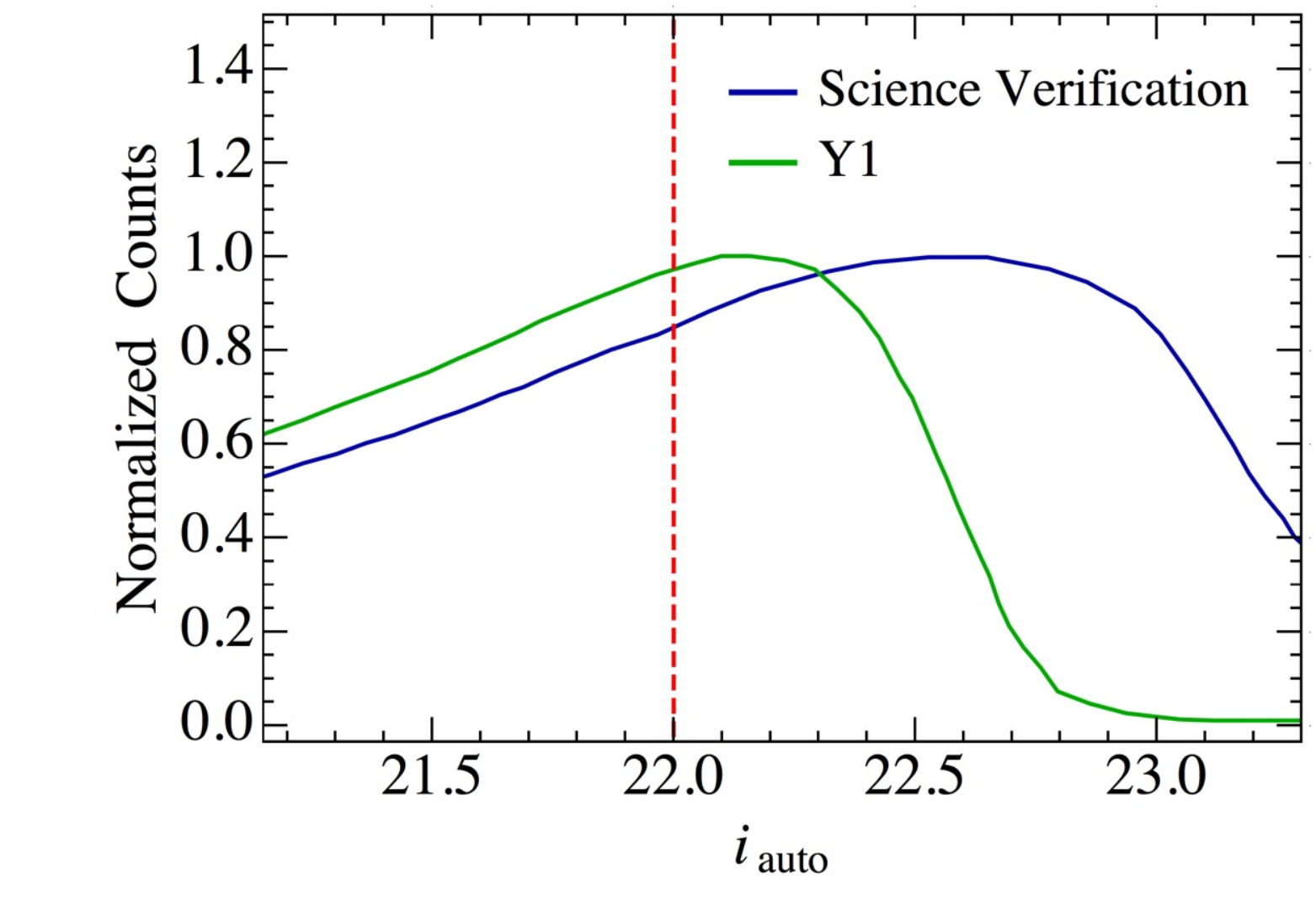}
\caption{Measurement of the trade off between area and number of
  objects as a function of magnitude limit and sample flux limit in Y1GOLD
  and SV. For a given $i_{\rm auto}$-band ``threshold'' value we select all
  regions which have a deeper limiting magnitude that this value  (10$\sigma$ depth
  limit $>$ ``threshold'') and count the galaxies brighter than the
  ``threshold'' value over those regions. These should be complete samples
  at each threshold value. Number counts are shown normalized
  to their maximum in the figure.}
\label{Fig:tradeoff}
\end{figure}

The overall flux-limit of the sample is set as
\begin{equation}
i_{\rm auto} < 22. 
\label{eq:faint}
\end{equation}
Additionally, we remove the most luminous objects by making the cut
$i_{\rm auto} > 17.5$ . \change{
The cut of Eq.~(\ref{eq:faint}) is chosen as a compromise
between}\change{ survey area, given that we need to achieve an homogeneous depth, and the number of galaxies in that area. 
For a given overall
flux limit of the galaxy sample (e.g. all galaxies with $i \le 22$) we
select the regions of the survey that are deeper than that 
limit (e.g. $i$-band 10$\sigma$ limit depth $> 22$) and mask
everything brighter. In this way that sample selection
should be complete over such
footprint. Clearly, for fainter selections more
objects are incorporated into the sample but the area of the survey reaching that
depth homogeneously is also smaller. Hence there is a compromise
between area and number of objects. In Fig. \ref{Fig:tradeoff} we
show the normalized counts as} \change{a function of the magnitude limit cut.
For comparison we include the same quantity in
Science Verification Data, which is deeper than Y1 but has much
smaller area, see \cite{2016MNRAS.455.4301C}.
We would like to select a sample and footprint that are at once
homogeneous and with the highest possible number of galaxies.
The curve shows a plateau in the range  $22 \lesssim i_{\rm
  auto} \lesssim 22.3$ where the
number counts is maximized, with variations of about $5\%$. But the figure does not account for 
photo-$z$ performance, which degrades rapidly for fainter objects
(particularly at high redshift) and is of key relevance for BAO
measurements, as shown below in Sec. \ref{sec:optimal-mag-color-cut}. Therefore we decided to stay at the
bright end of this range ($i_{\rm auto}=22$) as an overall flux limit
of the sample.}


Color outliers which are either unphysical or from special samples
(Solar System objects, high redshift quasars) are removed as well, to avoid extraneous photo-$z$ populations in the sample
(see Table \ref{Tab:sampledef}).

\subsection{Star-Galaxy Separation}
\label{sec:cuts2}

Removing stars from the galaxy sample is an essential step to avoid
the dampening of the BAO signal-to-noise \citep{2012MNRAS.419.1689C}
or the introduction of spurious power on large scales
\citep{Ross11}. Stellar contamination affects the broad shape of the
measurement and so we want to minimise it to be able to fit the BAO
template properly. However, it does not appreciably affect the
location of the BAO feature, so we do
not need to push for 100\% purity. Any residual contamination is then taken care of by using the weighting scheme detailed in Section \ref{sec:obs_syst}.

\begin{figure}
\begin{center}
\includegraphics[width=0.9\linewidth]{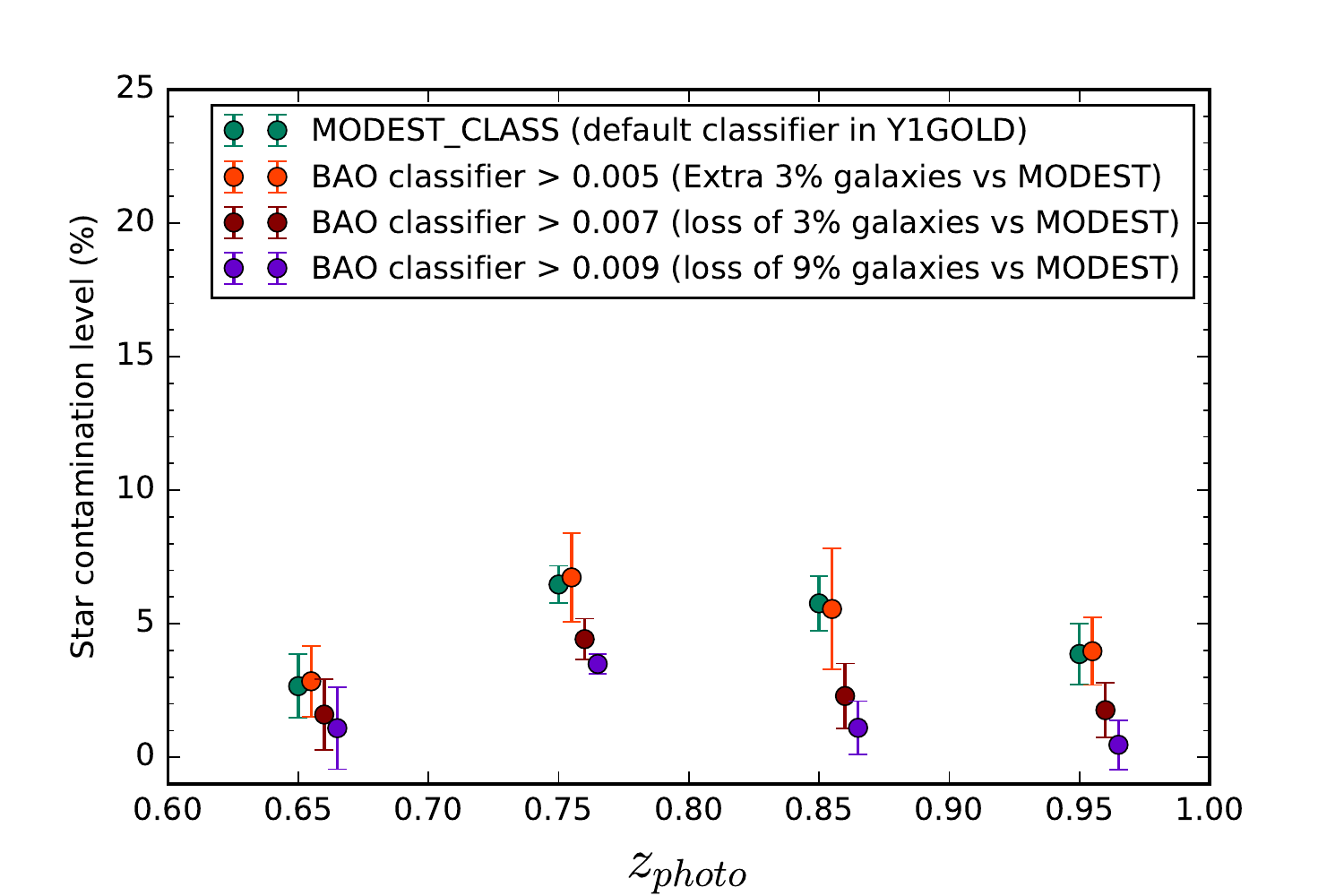}
\caption{Contamination of galaxy sample from stars as a function of redshift and star-galaxy separation threshold, as measured using galaxy density vs stellar density plots (from a pure stellar sample). The {\tt MODEST} classifier is defined in \citet{Y1Gold} as the default star galaxy classifier (based on ${\tt spread\_model}$ and ${\tt wavg\_spread\_model}$). `BAO classifier' stands for a cut in ${\tt spread\_model\_i} + (5.0/3.0){\tt spreaderr\_model\_i}$. A threshold of 0.007 provides an important decrease of contamination with a minor adjustment in the number of galaxies, which becomes significantly more severe at higher thresholds for a very similar purity. The redshift binning here uses $z_{\rm BPZ-AUTO}$.} 
\label{Fig:sgsamples}
\end{center}
\end{figure}

\begin{figure}
\begin{center}
\includegraphics[width=0.9\linewidth]{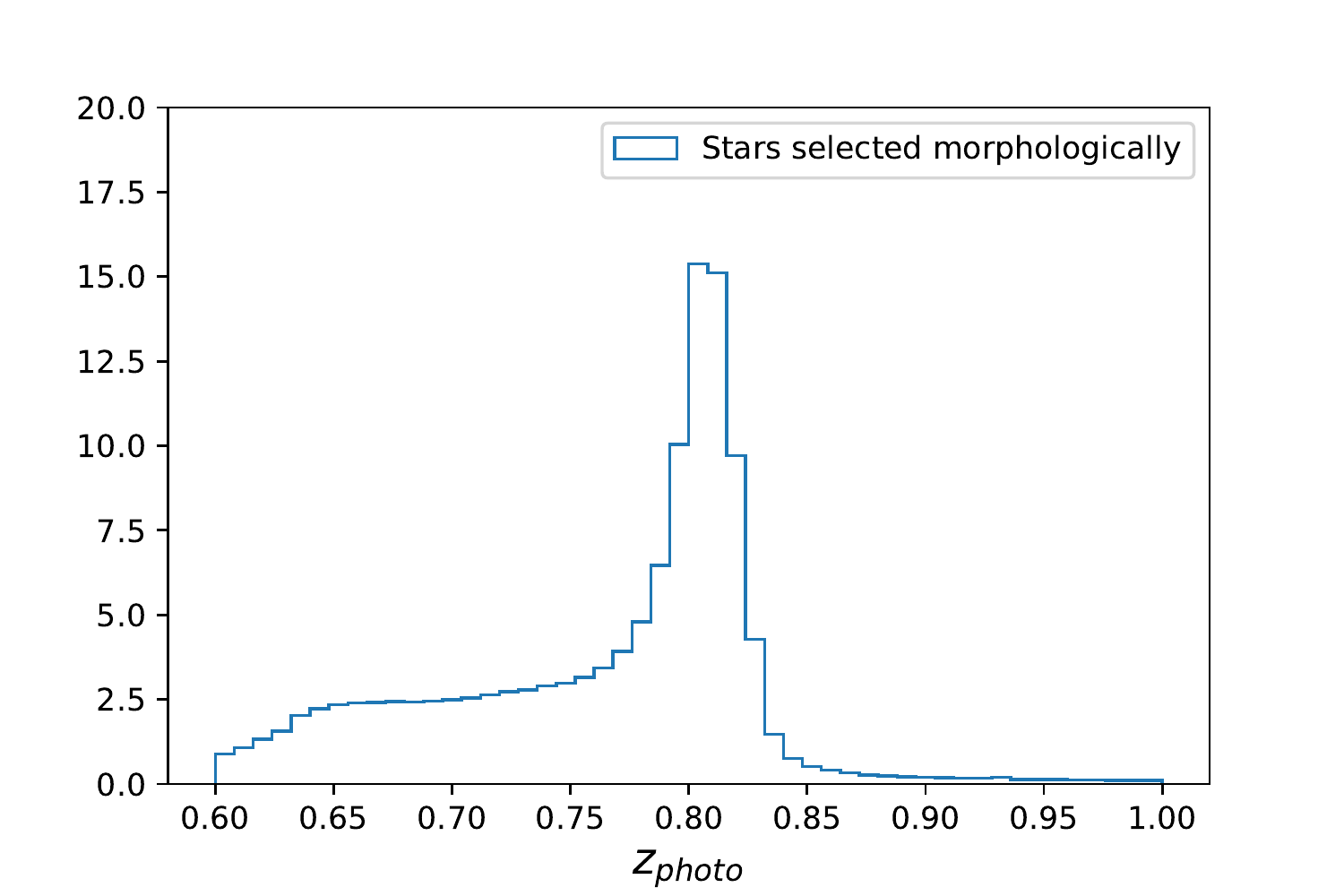}
\caption{Photometric redshift distribution of stars selected morphologically and passing the same cuts described in Table \ref{Tab:sampledef}.The redshift value $z_{phot}$ is the mean from the pdf of $z_{\rm BPZ-AUTO}$, which was used for the overall sample selection in Section \ref{sec:sample_selection}.} 
\label{Fig:stars_pz}
\end{center}
\end{figure}

In this work we have used the default star-galaxy classification
\change{scheme described in detail in \cite{sgsepy1}, see
  also \cite{Y1Gold},} which is based on the $i$-band coadd magnitude ${\tt spread\_model\_i}$ and its associated error ${\tt spreaderr\_model\_i}$,  from {\tt SExtractor}. This classifier was developed using as truth tables data from COSMOS \citep{2007ApJS..172..219L}, GOOD-S \citep{2004ApJ...600L..93G} and VVDS \citep{2005A&A...439..845L} overlapping Y1GOLD, and subsequently tested against CFHTLenS \citep{2013MNRAS.433.2545E}. The combination  ${\tt spread\_model\_i} + (5.0/3.0){\tt spreaderr\_model\_i} > 0.005$ is suggested for high-confidence galaxies as a baseline for Y1GOLD. A detailed follow up analysis of star-galaxy separation is given in \cite{sgsepy1}. Here instead we decided to modify slightly this proposed cut in order to increase the purity of the sample (from $95 \%$ to $97-98 \%$), at the cost of losing approximately $3\%$ of the objects, by making the following selection:
\begin{equation}
{\tt spread\_model\_i} + (5.0/3.0){\tt spreaderr\_model\_i} > 0.007.
\nonumber
\end{equation}

\begin{figure*}
\begin{center}
\includegraphics[trim = 0.5cm 12cm 0.5cm 11.5cm, width=0.99\linewidth]{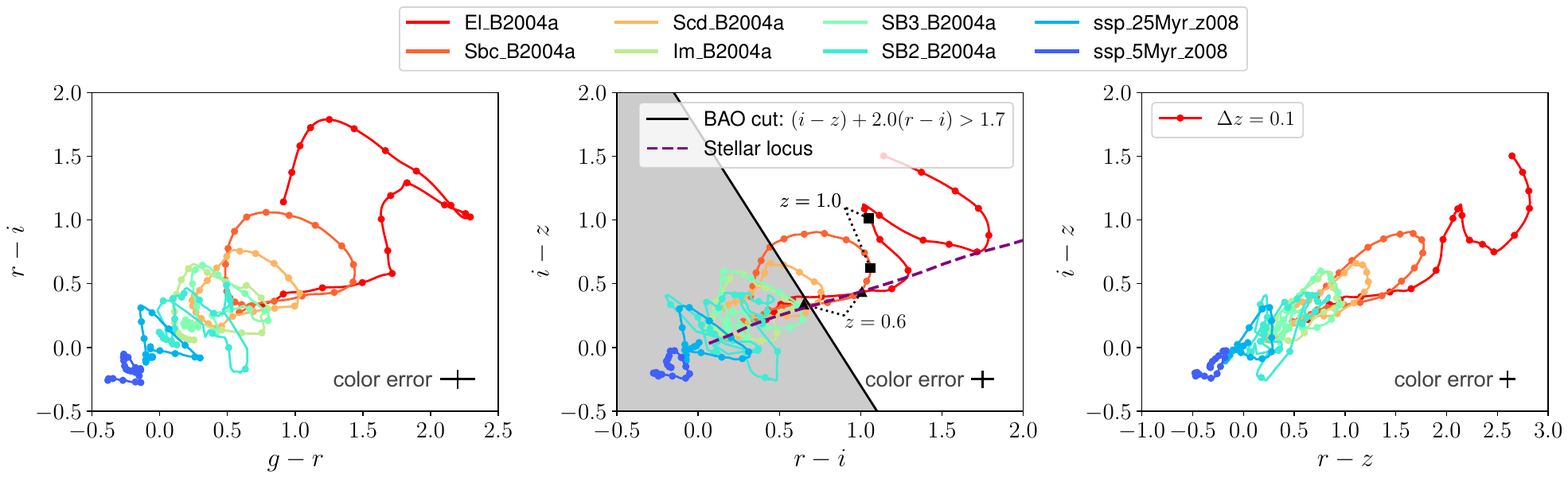}
\caption{Evolution of BPZ templates in color-color space. Each dot corresponds to a
  different redshift in steps of 0.1, ranging from $z=0.0$ to
  $z=2.0$. The shadowed region in the central panel is excluded from
  the sample. The black dots indicate the position of $z=0.6$
  (triangles), and $z=1.0$ (squares) for the two reddest
  templates. Also shown, for reference, is the stellar locus as a
  purple dashed line. \change{The inset crosses indicate an estimate of the
  error in the colors, arising from photometric errors, from a
  sub-sample of DES Y1 galaxies selected in
the range $21 < i_{\rm auto}<22$ (see text for more details)}.} 
\label{Fig:color-tracks}
\end{center}
\end{figure*}

In Fig. \ref{Fig:sgsamples} we show the estimated star sample
contamination for different thresholds of this cut, using the relation
between galaxy density and a map of stellar density built from Y1GOLD
(a methodology that is described in detail in section \ref{sec:obs_syst}). \change{The error bars displayed are the fitting errors obtained for the intercept when parametrizing the contamination level using a linear relationship between the galaxy density as a function of stellar density.}
Note that a threshold of $0.007$ reduces the contamination level to less than $5\%$ across the redshift range of interest. In Table \ref{tab:sample} we report a consistent or smaller level of stellar contamination, using a similar estimation, in the catalogues with MOF photometry, both for BPZ and DNF (see Sec. \ref{sec:obs_syst}). In Fig. \ref{Fig:color-tracks} we also include in the middle figure the track from the stellar locus, which showcases the reason why the first two redshift bins are more affected by stellar contamination, as it crosses the elliptical templates at these redshifts. To further illustrate this, in Fig. \ref{Fig:stars_pz} we show the distribution of the mean photometric redshifts for stars (selected using the criterion $|{\tt wavg\_spread\_model\_i}| < 0.002$, a more accurate variant of ${\tt spread\_model\_i}$ using single-epoch, suitable for moderate to bright magnitude ranges) showcasing how they will contaminate preferentially the second redshift bin, following the same trend as shown in Table \ref{tab:sample}.

\subsection{Selecting Red Luminous Galaxies}
\label{sec:cuts3}

Next we want to select from Y1GOLD a sample dominated by luminous red galaxies,
because their typical photo-$z$ estimates are more accurate than for the average galaxy population, thanks to the
4000 \AA\ Balmer break in their spectra. This feature makes redshift determination easier
even with broad-band photometry \citep{Pad05}. In addition we want our BAO sample to cover
redshifts larger than $0.6$ as there are already very precise BAO measurements
for $z < 0.6$, see e.g. \cite{2016MNRAS.457.1770C,2017MNRAS.464.1168R,2017MNRAS.464.3409B}.

We have tested that, while a very stringent selection can be done to
yield minimal photo-$z$ errors, e.g. with the redMaGiC algorithm \citep{2016MNRAS.461.1431R}, it
does not lead to optimal BAO constraints because the sample ends up being very
sparse, with $\sim 200,000$ galaxies in Y1GOLD at $z>0.6$ \citep{ElvinPoole17}. 
Instead we will follow an alternative path and apply a standard
selection in color-color space to isolate red galaxies at high
redshift, balancing photo-$z$ accuracy and number density
with a BAO figure-of-merit in mind.

In Figure ~\ref{Fig:color-tracks} we show the evolution in redshift of
the eight spectral templates used in BPZ, which includes one typical
red elliptical galaxy, two spirals and five blue irregulars/starbursts (color
coded) based on \cite{1980ApJS...43..393C} and
\cite{1996ApJ...467...38K}. We compute the expected observed DES broad-band magnitudes for these templates as a function of redshift and show them in different color-color combinations.The tracks are
evolved from $z=0$ to $z=2.0$ in steps of $0.1$ (marked with dots).
\change{We will use them to define cuts in color-color space intended to
isolate the red templates.} 

\change{In real data galaxy colors have an
uncertainty due to photometric errors, which effectively thicken those
tracks. In order to provide an estimate for this we computed the
errors in the colors for a sub-sample of Y1GOLD galaxies with $21 <
i_{\rm auto} < 22$ (the typical range of magnitudes that we explore
below to define the BAO sample). For each galaxy we estimate the color
error adding in quadrature the corresponding magnitude
errors\footnote{In turn computed as $m_{\rm err} = -2.5
  (Flux_{\rm err} / Flux)/\log(10)$}. The average error in each corresponding color is shown
with a cross at the bottom right inset label of the three panels of
Fig.~\ref{Fig:color-tracks}. Their values are 0.128, 0.073, 0.067,
0.076 for ({\it g-r, r-i, i-z, r-z}) respectively.}

\begin{figure}
\includegraphics[width=0.95\linewidth]{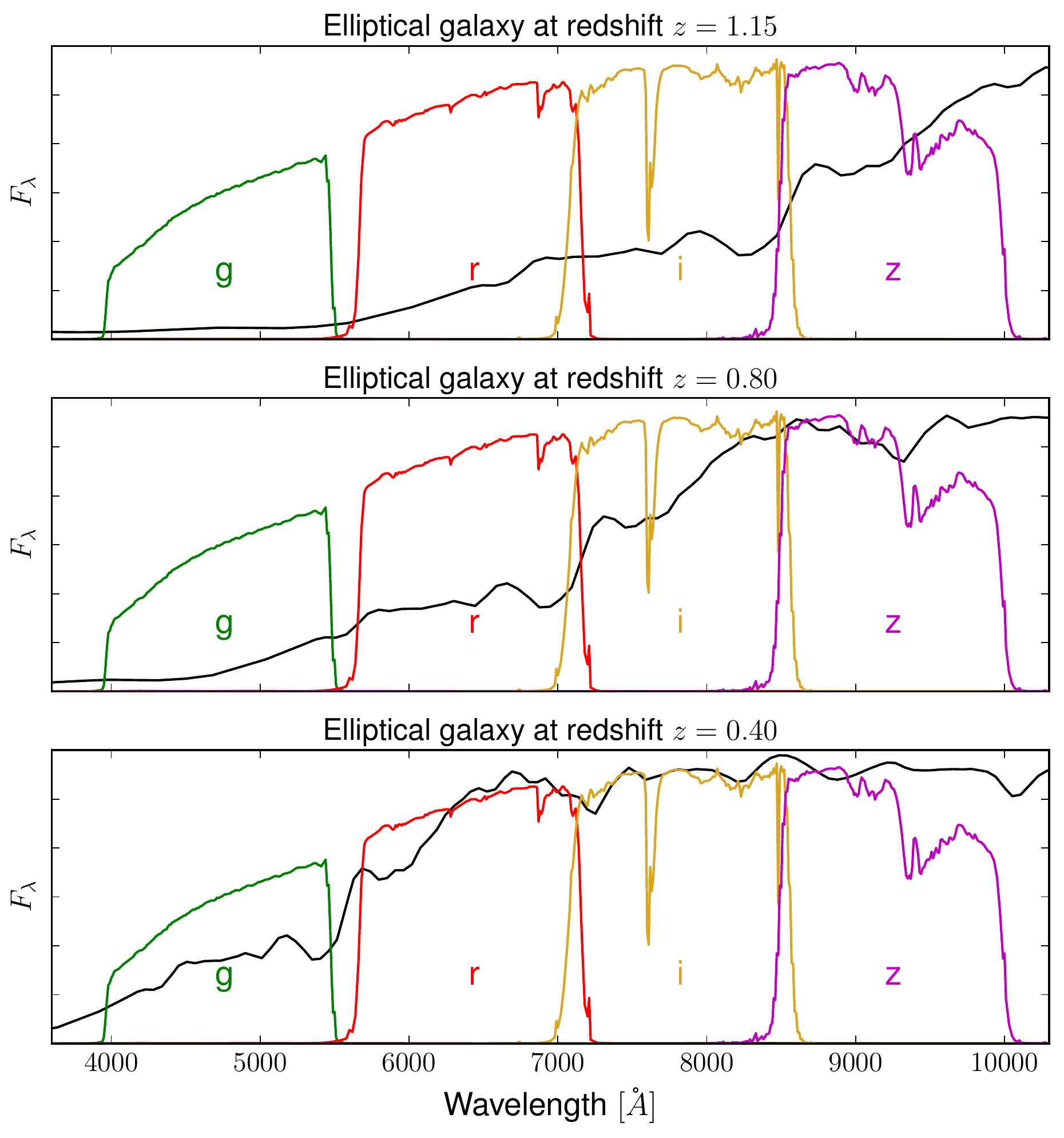}
\caption{Elliptical model spectrum used in template-based fitting code
  BPZ. Overplotted are the DES response filters g,r,i,z. The template
  has been redshifted to $z=0.4,\,0.8,\,1.15$, where the notable 4000
  \AA\,break crosses from $g\rightarrow r$, $r\rightarrow i$ and $i\rightarrow z$.   } 
\label{Fig:elliptical-des}
\end{figure}

\change{In addition,} a model of a red elliptical galaxy spectrum is shown in Figure~\ref{Fig:elliptical-des}, redshifted to $z=0.4,\,0.8,\,1.15$, where the notable 4000 \AA\,break crosses from $g\rightarrow r$, $r\rightarrow i$ and
$i\rightarrow z$. This suggests that for $z>0.6$ the strongest
evolution in color will be for $i-z$ and $r-i$, and hence we will
focus in these color combinations in what follows (\change{that moreover have the
  smallest error}). 

Note how the transition of the 4000 \AA\,break
from one band to another abruptly bends the color-color tracks in
Figure ~\ref{Fig:color-tracks}. However, this applies mainly to
elliptical templates, and recent star formation will dampen this
effect.

\subsection{Optimization of the color and magnitude cuts for BAO}
\label{sec:optimal-mag-color-cut}


\change{Optimizing the actual sample selection for the measurement of BAO in
imaging data is considerably different that doing so for
spectrospopic data. In the later case one basically needs to
maximize the area (or volume) provided that ${\bar n} P > 1$
(where ${\bar n}$ is the galaxy density and $P$ the power spectrum). For imaging
data the photometric redshift accuracy plays a vital role. Worse
photo-$z$ error degrades the signal as the galaxy radial separations are
smeared out (this also complicates the definition of survey
volume). In turn, the best photo-$z$'s are typically obtained for very
bright, and low density, samples. 
Therefore there is a non-trivial.
interplay to maximise BAO signal to noise. }

\change{In \methodXi\,we discussed in detail
how to fold in the photo-$z$ accuracy into an effective ${\bar n}_{\rm
  eff}$\footnote{Photometric redshift errors leads to ${\bar n}_{\rm eff} P < 1$ in
all cases explored.}. However computing ${\bar n}_{\rm eff}$ 
is cumbersome and as complicated as doing an actual
BAO forecasting. Therefore we decided to follow this later path and rely on
the Fisher matrix forecast formalism described in \cite{SeoEis07}. Provided with a concrete
set of color-magnitude cuts we measure in the data the
 number density and redshift uncertainty in several tomographic bins
 within 0.6 $\le$ photo-$z$ $\le$ 1.0, and assume a
 clustering amplitude. We then use the formulae from \cite{SeoEis07}
 to predict the precision that one can achieve with that set of galaxy
 data properties. We repeat this process for a different set of cuts until an optimal BAO
 distance error is achieved.}

\change{Through this process we fix the clustering amplitude, assuming 
a galaxy bias of $b=1.6$ for all calculations. This is the bias found in
\cite{2016MNRAS.455.4301C} for a flux limited sample ($i < 22.5$) at redshifts $z \sim
0.9$, selected from DES Science Verification (SV) data. Since
that redshift and magnitude are compatible with what we expect in this paper,
we consider $b=1.6$ a representative value. More precise measurements are expected for more biased samples, but the galaxy bias for any given sample is not known a priori and the redshift uncertainty and number density are the more dominant 
factors.}

\change{For illustrative purposes we show in Table~\ref{tab:variations} the variation
in BAO distance error achieved by changing the number density and
photo-$z$ accuracy away from those at the optimal cuts described below. We
also include the variation with survey area. As pointed before, BAO
distance errors are very sensitive to photo-$z$ accuracy.}

\begin{table}
\centering
\caption{Sensitivity of the forecasted BAO distance error to
  variations in density, photometric redshift errors and survey area.
  Note that these variations are considered individually, neglecting
  their correlations. Baseline values are those corresponding to the
  optimal cuts discussed in Sec.~\ref{sec:optimal-mag-color-cut}.}
\begin{tabular}{ll}
\hline
\hline
${\rm property\,variation}$  & ${\rm forecasted\,BAO\,distance\,error}$ \\
\hline
$10\%$ worse photo-$z$ & $8\%$ worse \\
$20\%$ worse photo-$z$ & $16\%$ worse \\
$10\%$ lower density & $3\%$ worse \\
$20\%$ lower density & $6\%$ worse\\
$10\%$ smaller area & $2.8\%$ worse \\
\hline
\hline
\end{tabular}
\label{tab:variations}
\end{table}

\subsubsection{Optimization of the color cut}
\label{sec:cuts3-color}

Thus, in order to maximize the signal-to-noise of the BAO \change{forecasted} measurement, a color cut is applied to the sample in the form,
\begin{equation}
(i_{\rm auto} - z_{\rm auto}) + a_{1} (r_{\rm auto} - i_{\rm auto}) > a_{2}. 
\label{equ:colorcut}
\end{equation}

The cut was chosen in this form following the discussion in Sec.~\ref{sec:cuts3}
(see Fig. ~\ref{Fig:color-tracks}), as it allows us to select more likely the 
reddest galaxies which are the ones with lower uncertainties in their photometric redshift determination and still present a high enough number density.

Samples were produced across a grid of $a_{1}$ and $a_{2}$ 
values, calculating the number of galaxies $N_{\rm gal}$ and a mean width
of the photo-$z$ distribution $\sigma_{z}/(1+z)$ for each sample, after splitting
the galaxy in tomographic bins. For BPZ we estimated $\sigma_z$
averaging in each tomographic bin the width of the individual redshifts
posterior distributions (PDFs) provided per galaxy. 

The BAO forecast using the algorithm of \cite{SeoEis07} is then run for the $N_{\rm gal}$ and 
$\sigma_{z}/(1+z)$ of each sample and final values of $a_{1}$ and $a_{2}$ 
are selected to minimise the forecasted BAO uncertainty, finding a balance between 
galaxy number density and redshift uncertainty. In order to give a
sense for the sensitivity of such process, we note there is a slight
degeneracy when increasing $a_{1}$ and $a_{2}$ simultaneously,
resulting in similar forecasted BAO uncertainties. However deviations from this degeneracy direction lead to significant degradation in the forecasted error. For example, doubling $a_{1}$ leads to a degradation of the forecasted error by approximately 0.01 (from $5\%$ to $6\%$ roughly). 
The values used in this analysis  are $a_{1} = 2.0$, $a_{2}  =
1.7$. Figure ~\ref{Fig:color-tracks} shows the color cut  in the central panel, where the shadowed region is excluded from the sample.

\subsubsection{Optimization of the magnitude cut}
\label{sec:cuts3-magnitude}

To further minimize the forecasted BAO uncertainty, an additional, redshift
dependent magnitude cut is applied to the sample as a second step. This applies a cut
to $i_{\rm auto}$ at low redshift which is stricter than the global
$i_{\rm auto} < 22$ cut (at lower redshift
the sample is sufficiently abundant that one can still select brighter
galaxies, with better photo-$z$, and still be sample variance
dominated). The cut is in the form, 
\begin{equation}
i_{\rm auto} < a_{3} + a_{4} z.
\end{equation}
As with the color cut in Eq.~\ref{equ:colorcut}, this is designed to
find a sample that balances redshift uncertainty with number density,
to minimise the forecasted BAO error. The BAO forecast error was
minimised at the values $a_{3} = 19$ and $a_{4} = 3$ and this cut was
applied to the sample. We find that the forecasted error improves by $\sim
15\%$ when introducing the redshift dependent flux limit as opposed to
a global $i_{\rm auto} < 22$ cut.

The final forecasted uncertainty on angular diameter distance combining all
the tomographic bins is $\sim 4.7\%$. Note that the discussion in this section
only has as a goal the definition of the sample. The real data
analysis with the sample defined here, and the final BAO error achieved, 
 will of course depend in many other variables that were not considered up to this point. Such as the
quality of photometric redshift errors, analysis and mitigation of systematics, use
of the full covariance and optimized BAO extraction methods. 

\change{Nonetheless we stress} that the forecasted error obtained in this section
matches \change{the one from the analysis of mock simulations, see e.g. \methodACF}, and is in fact
quite close to the final BAO error obtained in \main. In the following
sections we discuss the various components that will enter the real
data analysis, starting with the validation of photometric redsfhit
errors and the estimate of redshift distributions.

\section{Photometric Redshifts}
\label{sec:photozvalidation}

\begin{table}
\centering
\caption{Characteristics of the DES Y1 BAO sample, as a function of
 redshift. Results are shown for a selection of the sample in bins
 according to DNF photo-$z$ ($z_{\rm phot}$)  estimate in top of the table
 and  BPZ  in the bottom, both with MOF photometry. Here $\bar{z}=<z_{true}>$ is the
 mean true redshift, $\sigma_{68}$ and $W_{68}$ are the $68\%$ confidence widths of
$(z_{\rm phot}-z_{true})/(1+z_{true})$ and  $z_{true}$
respectively, all estimated from COSMOS-DES validation with SVC
correction, as detailed in Sec.~\ref{sec:photozvalidation} and
Fig.~\ref{Fig:N(z)}. $f_{\rm star}$ is the estimated stellar
contamination fraction, see Sec.~\ref{sec:obs_syst}}
\begin{tabular}{l@{\hspace{0.35cm}}c@{\hspace{0.35cm}}c@{\hspace{0.35cm}}c@{\hspace{0.35cm}}c@{\hspace{0.35cm}}c@{\hspace{0.35cm}}c}
\hline
\hline
${\rm DNF}$  & $N_{\rm gal}$ & $bias$ & $\bar{z}$ & $\sigma_{68}$ & $W_{68}$ & $f_{\rm star}$\\
\hline
$0.6- 0.7$ &386057 & 1.81 $\pm$ 0.05 & 0.652 & 0.023 & 0.047 & 0.004 \\
$0.7-0.8$ & 353789 & 1.77 $\pm$ 0.05 & 0.739 & 0.028 & 0.068 & 0.037 \\
$0.8-0.9$ & 330959 & 1.78 $\pm$ 0.05 & 0.844 & 0.029 & 0.060 & 0.012 \\
$0.9-1.0$ & 229395 & 2.05 $\pm$ 0.06 & 0.936 & 0.036 & 0.067 & 0.015 \\
\hline
\hline
${\rm BPZ}$ & $N_{\rm gal}$ & $bias$ & $\bar{z}$ & $\sigma_{68}$ & $W_{68}$ & $f_{\rm star}$\\
\hline
$0.6-0.7$ & 332242 & 1.90 $\pm$ 0.05 & 0.656 &0.027 & 0.049 & 0.018 \\
$0.7-0.8$ & 429366 & 1.79 $\pm$ 0.05 & 0.746 & 0.031 & 0.076 & 0.042 \\
$0.8-0.9$ & 380059 & 1.81 $\pm$ 0.06 & 0.866 & 0.034 & 0.060 & 0.015 \\
$0.9-1.0$ & 180560 & 2.05 $\pm$ 0.07 & 0.948 & 0.039 & 0.068 & 0.006 \\
\hline
\hline
\end{tabular}
\label{tab:sample}
\end{table}

The photometric redshifts used for redshift binning and transverse distance computations in our fiducial analyses are derived using the  Directional Neighborhood Fitting (DNF) algorithm  \citep{2016MNRAS.459.3078D}, which is trained with public
spectroscopic samples as detailed in \cite{2017arXiv170801532H}.
For comparison we also discuss below the Bayesian Photometric
Redshift (BPZ)  \citep{2000ApJ...536..571B} which we find slightly less performant in terms of the error with respect to ``true'' redshift values (see below). In both cases we use MOF photometry which provides $\sim 10-20\%$
more accurate photo-$z$ estimates with respect to the equivalent estimates using SExtractor \texttt{MAG\_AUTO} quantities from coadd
photometry. In this section we summarise the steps taken to arrive at these choices, based on a validation against data over the COSMOS field.

\begin{figure*}
\includegraphics[trim={0 0.7cm 0.7cm 0}, width=0.37\linewidth]{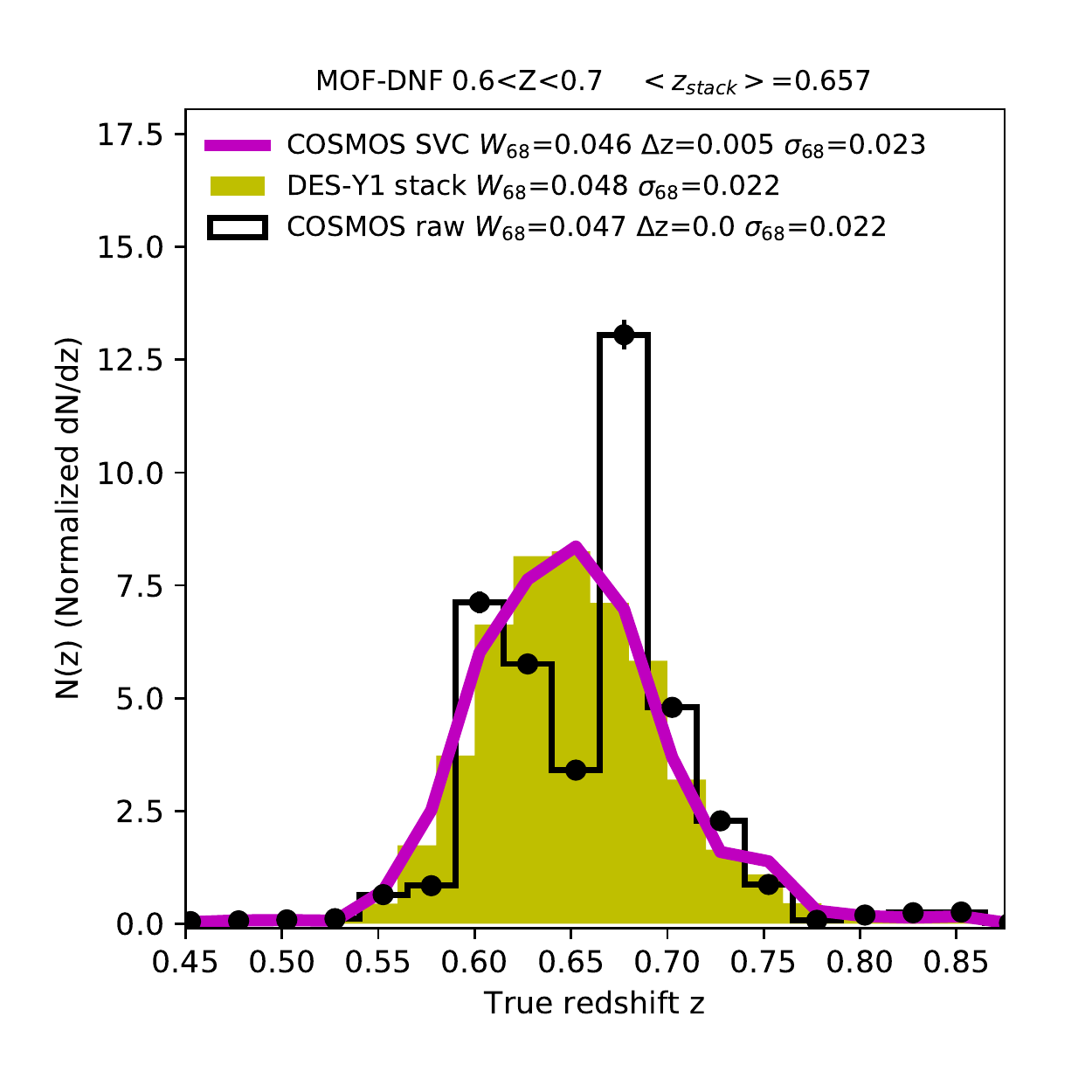}
\includegraphics[trim={0 0.7cm 0.7cm 0}, width=0.37\linewidth]{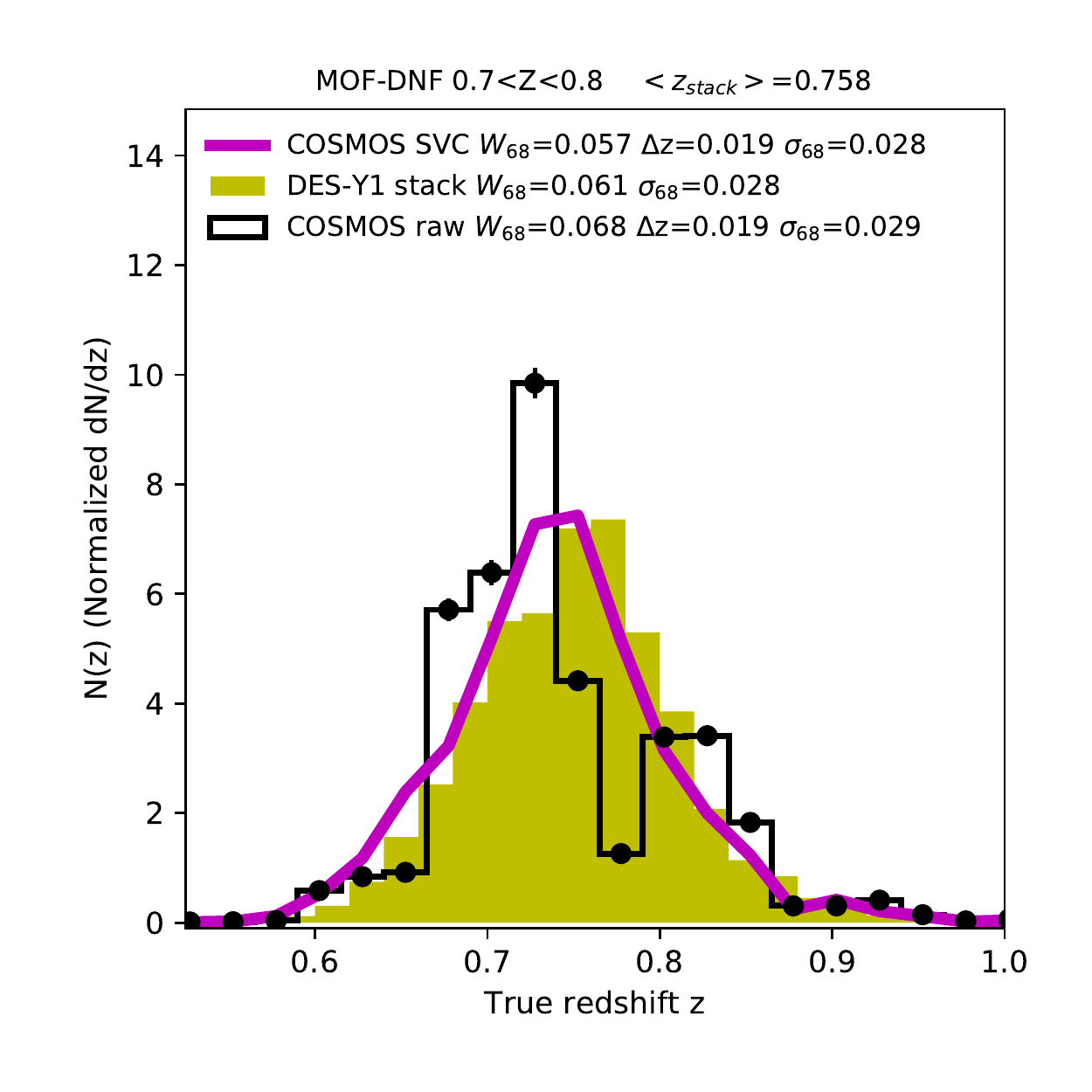} \\
\includegraphics[trim={0 0.7cm 0.7cm 0}, width=0.37\linewidth]{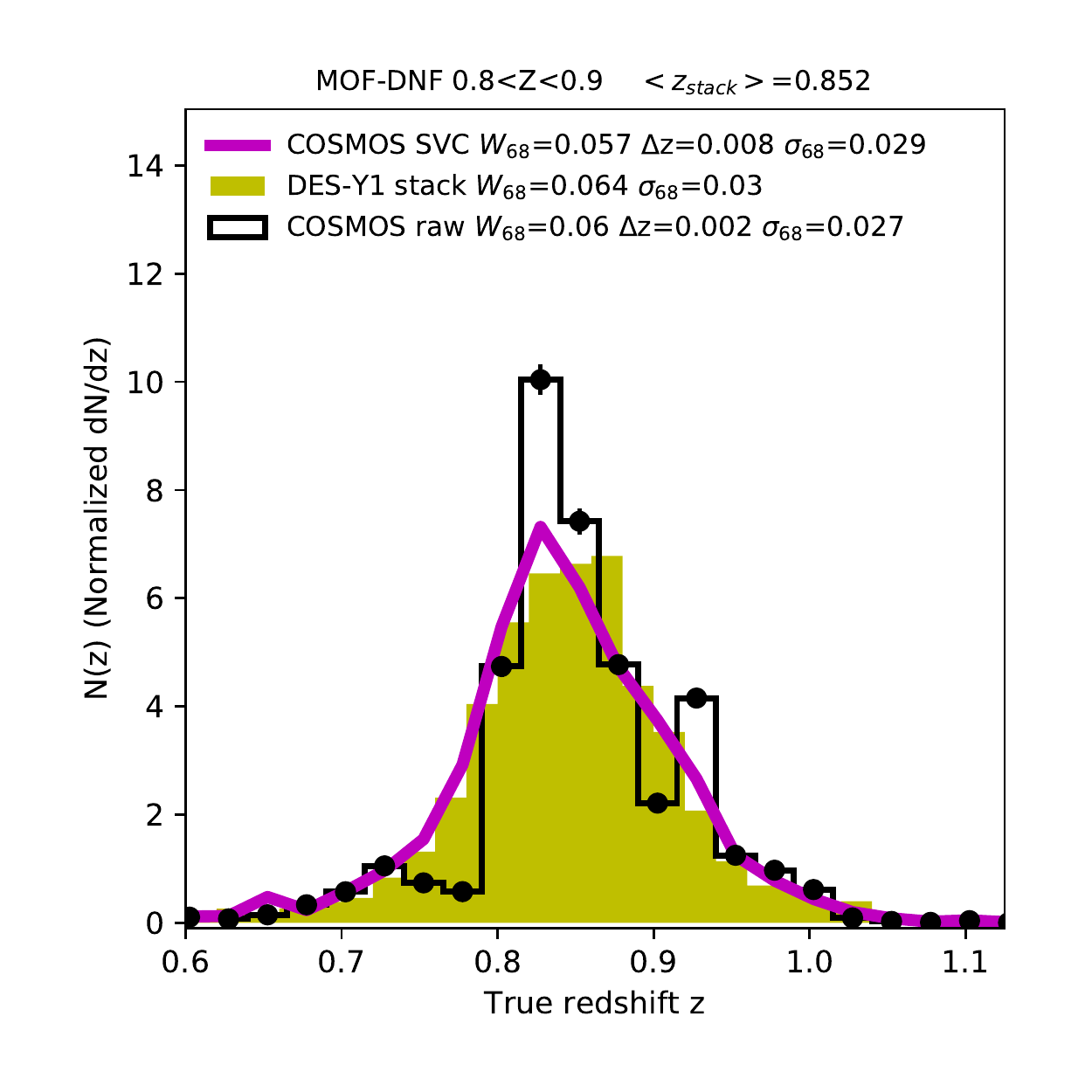}
\includegraphics[trim={0 0.7cm 0.7cm 0}, width=0.37\linewidth]{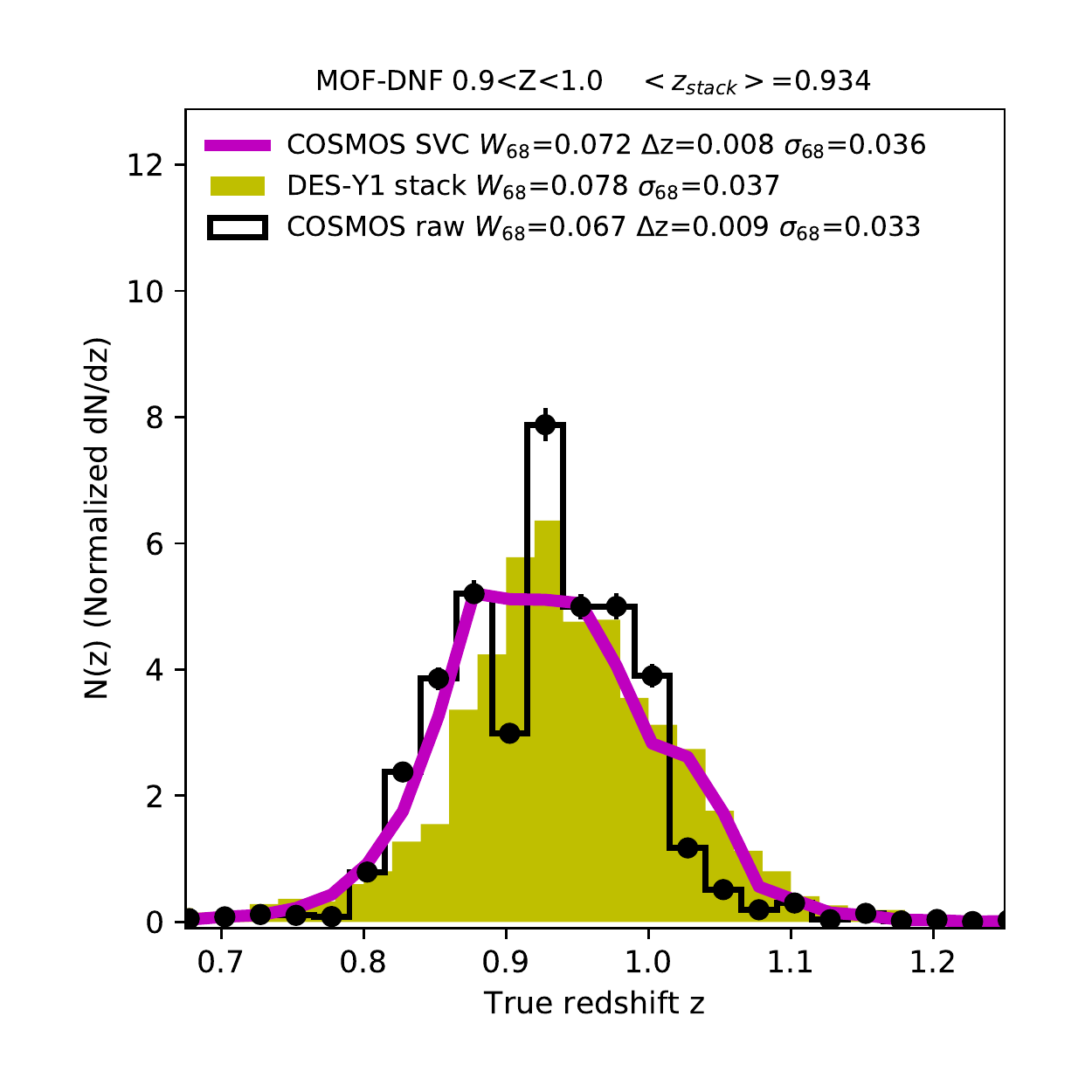}
\caption{Normalised redshift distributions for our different
  tomographic bins of DNF-MOF photo-$z$.  Stack $N(z)$ are shown for the
  full DES-Y1 BAO sample (yellow histograms).
The black histogram (with Poisson error bars) shows the raw 30-band photo-$z$ from
the COSMOS-DES validation sample. Magenta lines shows the same sample corrected by sample variance cancellation (SVC, see text),
which is our fiducial estimate. The labels show the values of
$W_{68}$, $\sigma_{68}$ and $\Delta z= <z_{stack}>-<z>$ and in each case, see also Table~\ref{tab:sample}.} 
\label{Fig:N(z)}
\end{figure*}

We recall that throughout this work we use the individual object's mean photo-$z$ from BPZ (not to be
confused with the mean value $\bar{z}=<z>$ of the sample) and the predicted value in the fitted hyper-plane from the DNF code, as our point estimate for galaxy redshifts.  As for the estimates of the  $N(z)$ from the photo-$z$ codes, for comparison with our fiducial choice based on the COSMOS narrow band $p(z)$, we will use the stacking of Monte Carlo realisations of the posterior redshift distributions $p(z)$ for the BPZ estimates, or the stacking from the nearest neighbour redshifts from the training sample, in the case of DNF (henceforth we'll call these \textit{stack} $N(z)$). Figure \ref{Fig:N(z)}  shows  the stack $N(z)$  (yellow histograms) in all 4 redshift bins for our fiducial DNF photo-$z$ analysis.

\subsection{COSMOS Validation}

As detailed in \photoz, we check the performance of each code by using 
redshifts in the COSMOS field (which are not part of the training set in the case of DNF), following the procedure outlined in \cite{2017arXiv170801532H}. These redshifts are either spectroscopic or accurate ($\sigma_{68}<0.01$) 30-band photo-$z$ estimates from \cite{2016ApJS..224...24L}. 
Both validation samples give consistent results in our
case because the samples under study are relatively bright.

The COSMOS field is not part of the DES survey. 
However a few select exposures were done by DECam which were processed by DESDM using the main survey pipeline. We call this sample DES-COSMOS. Because the COSMOS area is small
(2 square degrees) and DECam COSMOS images were deeper and not taken
as part of the main DES-Y1 Survey, we need to first resample the DES-COSMOS photometry
to make it representative of the full DES Y1 samples that we select in our
BAO analysis.  
Hence we add noise to the fluxes in the DES-COSMOS catalog to match the noise properties of the fluxes in the DES-Y1 BAO sample, this is what we refer to as resampled photometry.  Then for each galaxy in the DES-Y1 BAO sample, we select
the galaxy in DES-COSMOS whose resampled flux returns a minimum $\chi^2$ when compared to the DES-Y1 BAO flux
(the $\chi^2$ combines all bands, $g,r,i$ and $z$).
This is done for every galaxy in the DES-Y1 BAO sample to make up the `COSMOS-Validation' catalog, which by construction has colors matching those in the DES-Y1 BAO sample. The ``true'' redshift is retrieved from the spectroscopic/30-band photo-$z$ of this match. 

We then run the DNF photo-$z$ code over the COSMOS-Validation catalog to select 4 redshift bin samples in the same way as we did for the full DES-Y1 BAO sample. We use the ``true'' redshifts from the COSMOS-Validation catalogs to estimate the  $N(z)$ in each redshift bin by normalising the histogram of these true
redshifts. 

Results are shown as histograms in Figure \ref{Fig:N(z)}, which are compared to the stack $N(z)$ from the photo-$z$ code, for reference.  The black histograms show large fluctuations which are caused by real individual large scale structures in the COSMOS field. This can be
seen by visual inspection of the maps. This sampling variance comes from the relatively small size of the COSMOS
validation region. There is also a shot-noise component, indicated by the error bars over the black dots, but it is smaller. In the next section, we briefly describe the methodology to correct for this to be able to make use of this validation sample effectively.

\subsection{Sample variance correction}

As detailed in \photoz\, we apply a sampling variance correction (SVC)
to the data and test this method 
with the Halogen mocks described in \mocks. 
In what follows we provide a summary of such process and its main results. 

We use the VIPERS catalog \citep{2016arXiv161107048S}, which spans 24
square degrees to $i < 22.5$, 
to estimate the sampling variance effects in the above COSMOS validation. After correcting
VIPERS for target, color and spectroscopic incompleteness we select
galaxies in a similar way as done in section \ref{sec:sample_selection}. 
We then use the VIPERS redshifts to estimate  the true $N(z)$ distribution of the parent DES-COSMOS
sample (before we select in photometric redshifts). The ratio of the
$N(z)$ in the DES-COSMOS sample to the one in VIPERS gives a sample
variance correction that needs to be applied to the $N(z)$ in each of
the tomographic bins.

Figure \ref{Fig:N(z)} shows the SVC-corrected version of the raw
COSMOS catalog in magenta. As shown in
 this figure the resulting distribution is much smoother than the
 original raw measurements (black histograms). 
This by itself indicates that
 SVC is working well. Tests in simulations show that this SVC method
is unbiased and reduces the errors in the mean and variance of the $N(z)$ distribution by 
up to a factor of two. Similar results are found for different binnings in redshift.

\change{Notably, the distributions obtained from the {\it stacked N(z)} and the ones from COSMOS
SVC match well overall, although some discrepancies can be seen, e.g. for the
second and fourth bin. More quantitative statements are provided
below, but in \main\,(Table 5, entry denoted ``$w(\theta)$ $z$ uncal'') we show these have no impact in our cosmological results.
The difference in angular diameter distance measurements when using either of these two sets of redshift distributions 
is less than ∼ $\sim 0.25 \sigma$}. 


\subsection{Photo-$z$ validation results}

In Table \ref{tab:sample} we show the values of $\sigma_{68}$, which corresponds to
the $68\%$ interval of values in the distribution of
$(z_{photo}-z_{true})/(1+z_{true})$ around its median value, where $z_{photo}$ is the photo-$z$
 from DNF ($z_{mean}$ above) and $z_{true}$ is the redshift from the COSMOS validation
sample corrected by SVC. We also show $W_{68}$ and $\bar{z}$ which are the 68\% interval
and mean redshift in the $z_{true}$ distribution for each redshift
bin. The corresponding values for the $stack$ $N(z)$ and raw $N(z)$ are also
shown in the labels of  Figure \ref{Fig:N(z)}. $\Delta z$ in the
label inset shows the difference $\Delta z= <z_{stack}>-<z>$, where
$<z_{stack}>$ is the mean stack  redshifts for DES-Y1,  shown in the
top label.

\change{We have performed an extensive a comparison of the quantities shown
in Table \ref{tab:sample} computed with different validations sets:
DES-COSMOS with and without SVC, using $N(z)$ from DNF stacks, using
the COSMOS subsample with spectroscopic redshifts (as opposed to that 
with 30-band photo-$z$). We have also compared these $N(z)$ to the one
predicted by subset galaxies that have spectra within the BAO sample
over full DES-Y1 footprint. Furthermore we have performed a validation
using a larger spectroscopic sample in the VIPERS/W4 field ($\sim
4$ square degrees) which was}
\change{observed in DESY1 and is completely independent from the COSMOS
validation} \footnote{\change{The completeness of the VIPERS sample depends on
  galaxy type and has a color preseleccion to exclude galaxies at $z <
  0.5$. We have included all the suggested incompleteness factors \citep{2016arXiv161107048S}, but
  nonetheless have decided to use COSMOS-SVC as our fiducial validation
  set to avoid potential residuals.}}. \change{The results from these different
validation sets is that the means of the redshift distributions
$\langle z \rangle$ (w.r.t
to the mean using the {\it stack} $N(z)$) are always within 0.01 except for the
second tomographic bin where differences are $< 0.02$ (see also labels of Figure \ref{Fig:N(z)}). The values of  $W_{68}$
are always within $0.01$ as well, for all bins. This means that the
differences in $W_{68}$ are within $15\%-20\%$ (depending on redshift)
and $\langle z \rangle$ is within $1\%$ ($2\%$ for the bin $[0.7-0.8]$). 
In Sec. 4.3 of \methodACF\, we investigate the impact in derived BAO angular diameter
distances from systematic errors in the mean and variance of the underlying
redshift distributions. The most important quantity is the mean of
$dn/dz$. The level of shifts discussed above would induce about
$0.8\%$ systematic error in $\theta_{\rm BAO}$, while $20\%$ in the variance would have no
impact. These are small compared to the statistical errors, see \main.
The validation errors and biases in $\langle z \rangle$, $\sigma_{68}$
and $W_{68}$ were also studied
and we anticipate that they are subdominant for the BAO analysis, which instead is dominated by the
limited size of the DES Y1 footprint. These results will be presented
more extensively in \photoz.}


We also include in that work a  comparison with BPZ photo-$z$ (see also Table \ref{tab:sample})
 and results for different photo-$z$ with coadd photometry. The values of  $W_{68}$  and
 $\sigma_{68}$ are always  smaller (by 10-20\%)  for DNF
 with MOF photometry, which is therefore used as our fiducial photo-$z$
 sample. 


We finish the section by stressing that the fiducial $N(z)$ used in
the main BAO analysis are the ones from DES-COSMOS with
 SVC (magenta lines in  Figure \ref{Fig:N(z)}).

\section{Angular Mask}
\label{sec:mask}

We build our mask as a combination of thresholds/constraints on basic survey observation properties, conditions due to our particular sample selection, and restrictions to avoid potential clustering systematics. In summary,

\begin{itemize}
{\item We start by combining the Y1GOLD {\tt Footprint} and {\tt Bad regions} mask, both of which are described in \cite{Y1Gold}. The {\tt Footprint} mask imposes minimum total exposure times, valid stellar locus regression\footnote{This is a complementary calibration technique used for the construction of Y1GOLD making use of the distinct color locus occupied by stars to perform relative additional calibration between bands.} (SLR) calibration solutions and basic coverage fractions. The {\tt Bad Regions} mask removes at different levels various catalog artifacts, regions around bright stars and large foreground objects. In particular, for the later we remove everything with flag bit $>$ 2 in Table 5 of \cite{Y1Gold}, corresponding to regions around bright starts in the 2MASS catalogue \citep{2006AJ....131.1163S}.} \\ 

{\item We introduce coordinate cuts to select only the wide area parts
  of the surveys, namely those overlapping SPT (roughly with $300 < RA
  ({\rm deg}) < 99.6$ and $-40 < DEC ({\rm deg}) < -60$) and S82 (with
  $317.5 < RA ({\rm deg}) < 360$ and $-1.76 < DEC ({\rm deg}) <
  1.79$). This removes small and disjoint regions which are part of
  the Supernova survey and two auxiliary fields used for photo-$z$
  calibration and star-galaxy separation tests (COSMOS and VVDS-14h),
  which do not contribute to our clustering signal at BAO
  scales \change{(they total 30 deg$^2$)}.} \\

{\item Pixelized maps of the survey coverage fraction were created at
  a {\sc Healpix} resolution of N$_{\rm side}$ = 4096 (area = 0.73
  arcmin$^2$) by calculating the fraction of high resolution subpixels
  (N$_{\rm side}$ = 32768, area = 0.01 arcmin$^2$) that were contained
  within the original {\tt mangle} mask (see \cite{Y1Gold} for a
  description of the later). Since our color selection requires
  observations in all four $griz$ bands we use the coverage maps to
  enforce that all pixels considered, at resolution 4096, show at
  least $80\%$ coverage in each band \change{(this removes 70.7 deg$^2$
    with respect to the case where no miminum coverage is required)}. Furthermore we then use the minimum coverage across all four bands to down-weight the given pixel when generating random distributions, see Sec.~\ref{sec:two-point}.} \\

{\item In order to match the global magnitude cut of the sample and ensure it is complete across our analysis footprint, we select regions with $10\sigma$ limiting depth of $i_{\rm auto} > 22$, where the depths are calculated according to the procedure presented in \cite{Y1Gold}.} \\

{\item Since we want to reliably impose the color cut defined  in
  Eq.~(\ref{equ:colorcut}) and Table \ref{tab:sample},  we consider
  only areas with limiting depth in the corresponding bands large
  enough to measure it. Given that we are already imposing $i_{\rm
    auto}$ depth greater than 22, the new condition implies keeping
  only the regions with $10\sigma$ limiting magnitudes $(2\,r_{\rm
    auto} - z_{\rm auto}) < 23.7$, or equivalently those with $z_{\rm
    auto} > 2\,r_{\rm auto}-23.7$. \change{This removes an additional 53.8 deg$^2$.} }\\

{\item As a result of our analysis of observational systematics in Sec.~\ref{sec:obs_syst}, we identify that galaxy number density in regions of high $z$-band seeing shows an anomalous behaviour. To isolate this out we remove areas with $z$-band seeing greater than 1 arc-second \change{(this amounts to 71 deg$^2$, or $5\%$ of the footprint)}.} \\

{\item Lastly we also remove a \change{patch of $18\,{\rm deg}^2$} over which the airmass computation was corrupted.}  \\
\end{itemize}
The resulting footprint occupies 1336 deg$^2$ and is shown in Fig. \ref{Fig:footprint}.

\section{Mitigation of observational systematic effects}
\label{sec:obs_syst}

We have tested for observational systematics in a manner similar to \cite{ElvinPoole17}, which builds upon work in DES Science Verification Data \citep{2016MNRAS.455.4301C} and other surveys (e.g. \cite{Ross11,2012ApJ...761...14H}). 

Generically, we test the dependence of the galaxy density against survey properties (SPs). We expect there to be no dependence if SPs do not introduce 
density fluctuations in our sample beyond those already accounted for by the masking process. We have used the same set of SP maps as in \cite{ElvinPoole17}, namely :  
\begin{itemize}
{\item 10$\sigma$ limiting depth in band }
{\item full width half maximum of point sources (``seeing'')}
{\item total exposure time }
{\item total sky brightness,}
{\item atmospheric airmass,} 
\end{itemize}
all of them in each of the four bands $g r i z$, in addition to Galactic extinction and stellar contamination (refer to \cite{ElvinPoole17} for a detailed explanation on how the stellar density map is constructed from Y1GOLD data). We find that the relevant systematics are stellar density, PSF FWHM, and the image depth. We outline the tests that reveal this and how we apply weights to counter their effect in what follows.

\begin{figure}
\begin{center}
\includegraphics[width=84mm]{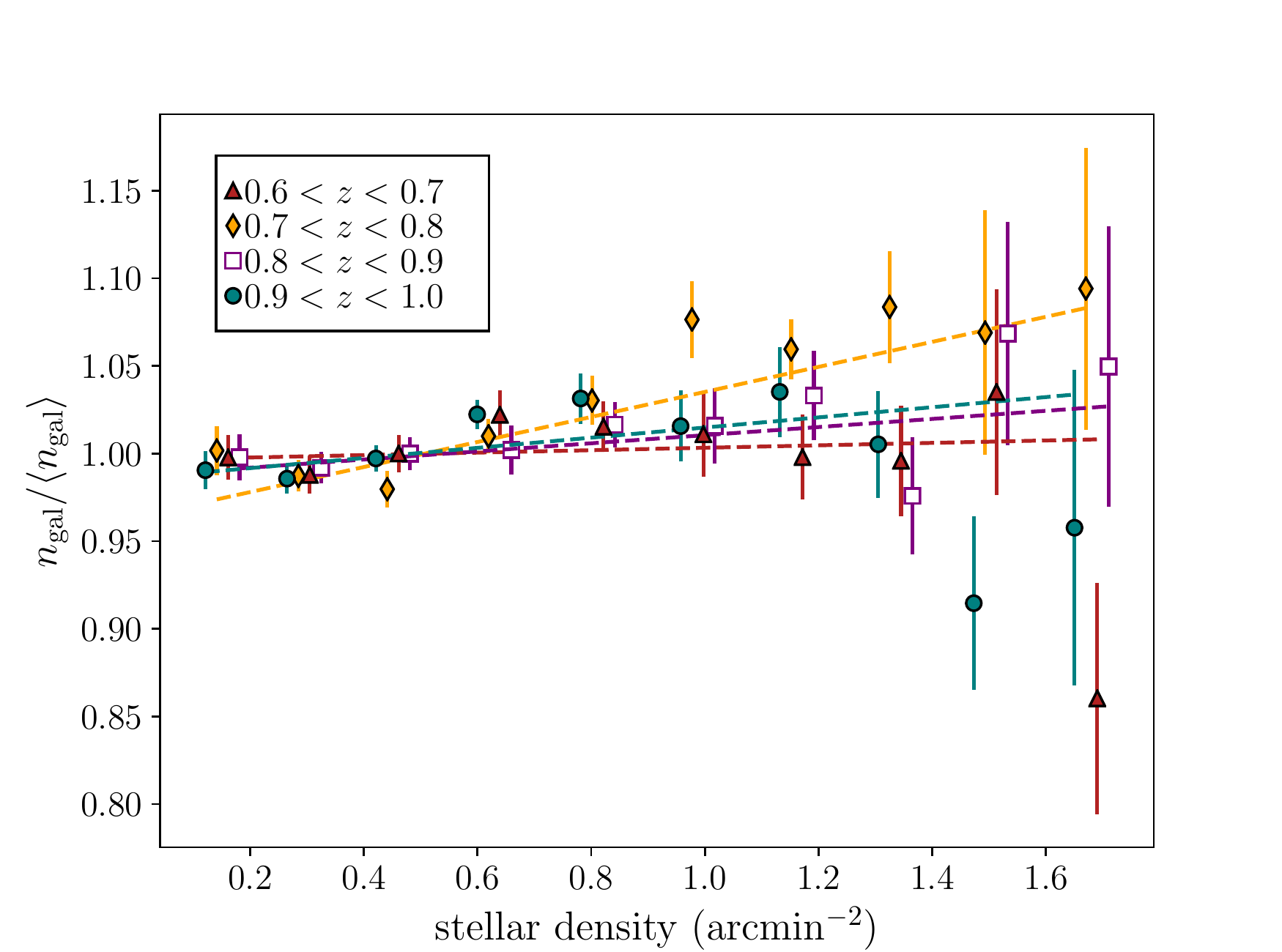}
\includegraphics[width=84mm]{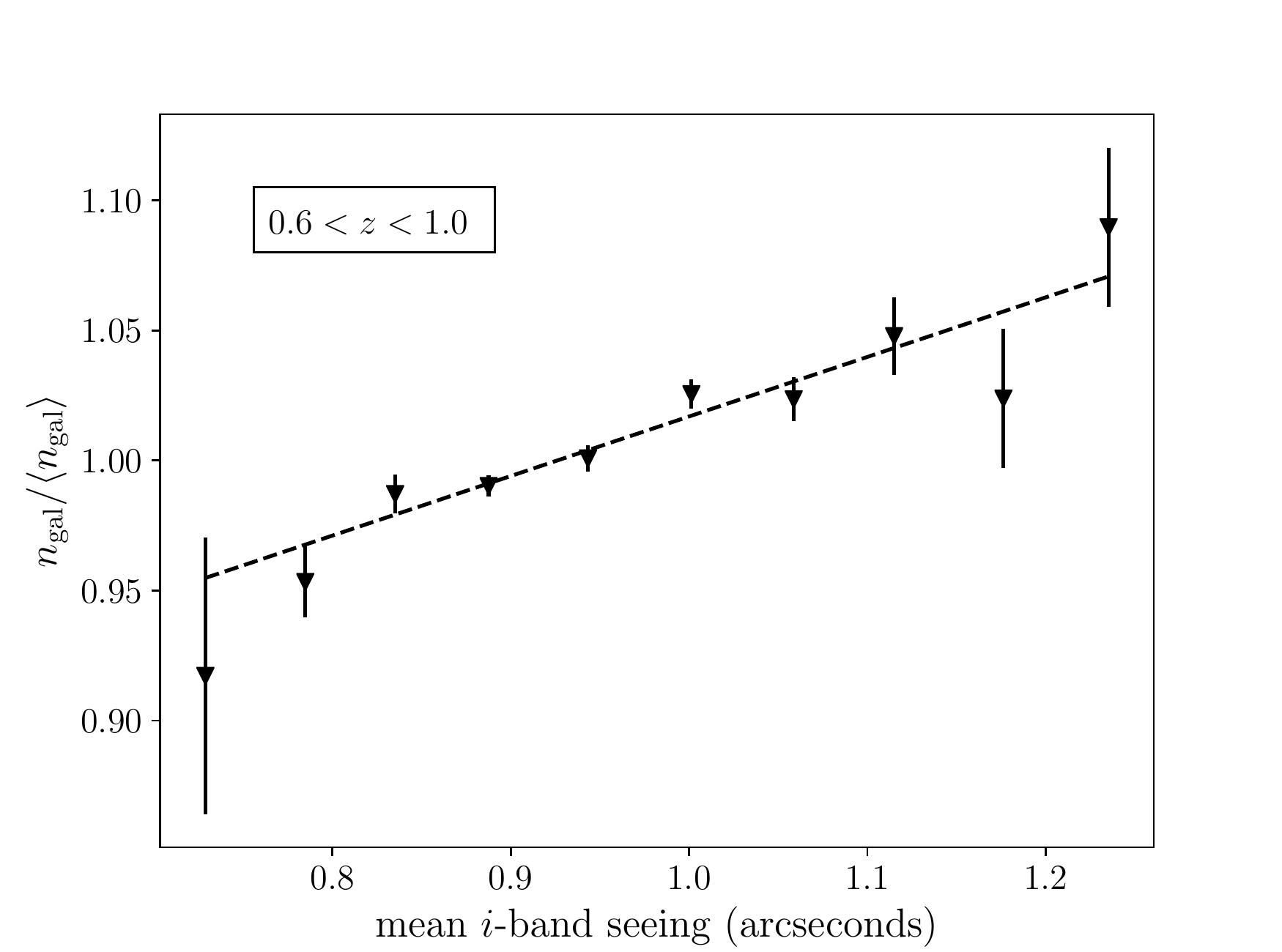}
\includegraphics[width=84mm]{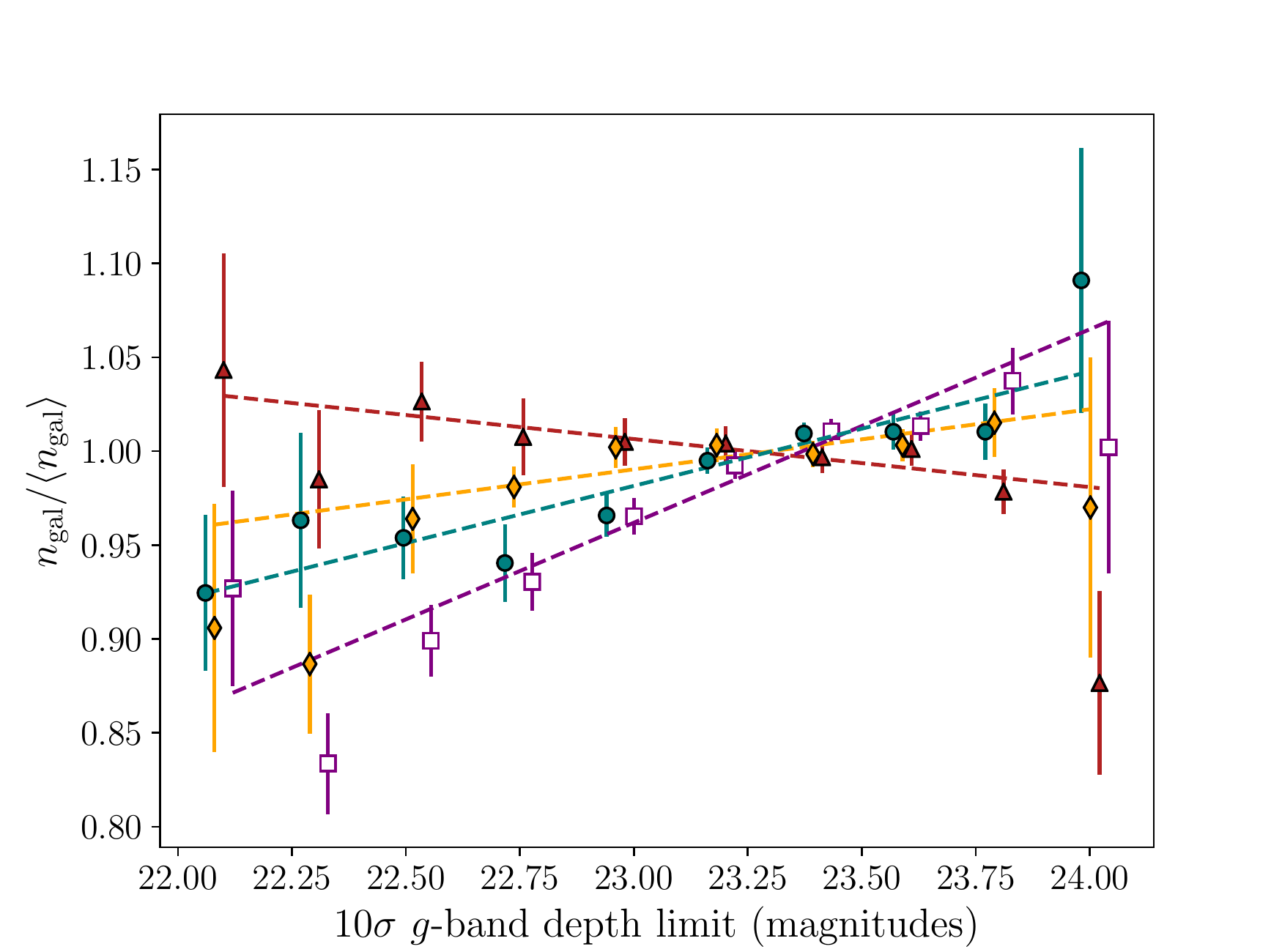}
\caption{The galaxy density vs. potential systematic relationship used to define weights that we apply to clustering measurements. {\it Top panel:} The galaxy density versus stellar density in four photometric redshift bins. The linear fits are used to determine the stellar contamination. The $\chi^2$ values for the fits are 9.7, 10.0, 3.5, and 14.3 (8 degrees of freedom). {\it Middle panel}: The galaxy density versus the mean $i$-band seeing for our full sample. The inverse linear fit is used to define weights applied to clustering measurements. The $\chi^2$ is 7.7 (8 degrees of freedom) and the coefficients are 0.788 and 0.0618. {\it Bottom panel}: The galaxy density versus $g$-band depth in four photometric redshift bins. The coefficients are interpolated as a function of redshift and used to define weights to be used in the clustering measurements. The $\chi^2$ values for the fits, given 8 degrees of freedom, are 7.7, 8.9, 12.7, and 6.1. The slopes are (-0.0256, 0.0320, 0.103, 0.0609).
  }
\label{fig:sysplots}
\end{center}
\end{figure}

We found the most important systematic effect, in terms of its impact on the measured clustering, to be the stellar density. In the top panel of Fig.~\ref{fig:sysplots} we find positive trends when comparing the number density of our `galaxy' sample as a function of the stellar number density ($n_{\rm star}$). Our interpretation is that there are stars in our sample. Assuming these contaminating stars follow the same spatial distribution as the stars we use to create our stellar density map, this stellar contamination will produce a linear relationship between the density of our galaxy sample and the stellar density. In this scenario, the value of the best-fit trend where the number density of stars, $n_{\rm star}$, is 0 is then the purity of the sample. We find the results are indeed consistent with a linear relationship, as illustrated in the top panel of Fig. \ref{fig:sysplots}. The stellar contamination, $f_{\rm star}$, that can be determined from these plots is listed in Table \ref{tab:sample}. The stellar contamination varies significantly with redshift, as expected given the proximity of the stellar locus to the red sequence as a function of redshift. Thus, we measure the stellar contamination in $\Delta_z = 0.05$ bin widths and use a cubic spline interpolation in order to obtain the stellar contamination at any given redshift. This allows us to assign a weight to each galaxy given by,
\begin{equation}
w(f_{\rm star}(z)) = \left((1-f_{\rm star}(z))+n_{\rm star}f_{\rm star}(z)/\langle n_{\rm star}\rangle\right)^{-1},
\end{equation}
where $n_{\rm star}$ is the stellar density that depends on angular location and $\langle n_{\rm star}\rangle$ is the mean stellar density over the DES-Y1 footprint.

Note that we repeat the fitting procedure for each photo-$z$ catalogue, hence redshift here means either $z_{\rm DNF-MOF}$ or $z_{\rm BPZ-MOF}$. From Fig.~\ref{fig:sysplots} it seems that the measurements are a bit noisy. However this procedure helps us resolve the peak in the stellar contamination of five per cent at $z\sim 0.78$. The uncertainty on each fit is $\sim 0.01$, which is consistent with the scatter we find in the values of $f_{\rm star}$ per bin. The spline simply interpolates between the best-fit values.

We also add weights based on fits against relationships with the mean
$i$-band PSF FWHM (seeing, which we denote as $s_i$) and the $g$-band
depth ($d_g$). For the seeing, we do not find a strong dependence on redshift and thus use the full sample to define the seeing dependent weight
\begin{equation}
w(s_i) = \left(A_s + B_ss_i\right)^{-1},
\end{equation}
where $A_s$ and $B_s$ are simply the intercept and slope of the best-fit linear relationship, shown in the middle panel of Fig. \ref{fig:sysplots}. The coefficients we use are $A_i=0.782$ and $B_i=0.0625$. For the $g$-band depth, we fit linear relationships in redshift bins $\Delta_z = 0.1$ and again use a cubic spline interpolation in order to obtain a weight at any redshift
\begin{equation}
w(d_g,z) = \left(C(z)+d_g(1-C(z))/\langle d_g\rangle\right)^{-1},
\end{equation}
where $C(z)$ is the interpolated result for the value of the linear-fit where $d_g = 0$. The relationships as a function of redshift and the linear best-fit models are shown in the bottom panel of Fig. \ref{fig:sysplots}. The total systematic weight, $w_{\rm sys}$, is thus multiplication of the three weights
\begin{equation}
w_{\rm sys} = w(f_{\rm star}(z))w(s_i)w(d_g,z).
\end{equation}

 \change{The dependencies we find are purely empirical as we
lack any more fundamental understanding for how these correlations
develop. They must result from the complicated intersection of our
color/magnitude selection and the photometric redshift algorithm, that
are not perfectly captured by our mask. 
Besides the relations with different observing properties (airmass,
seeing, dust, exposure time) are also very correlated what makes
physical interpretation very complicated.}

In the following section, we test the impact of these weights on the measured clustering, and determine their total potential impact. In \main\,, we show that the weights have minimal impact on the BAO scale measurements and that our treatment is thus sufficient for such measurements. Our treatment is not as comprehensive as \cite{ElvinPoole17}, and thus further study might be required when using the sample defined here for non-BAO applications.

\section{Two-point clustering}
\label{sec:two-point}

In this section we describe the basic two-point clustering properties of the samples previously defined. We concentrate on
large-scales where the BAO signal resides, and the sample using $z_{\rm
  DNF-MOF}$ photometric redshifts which is the default one used in \main. 

We compute the angular correlation function
$w(\theta)$ of the sample, split into four redshift bins, using the
standard Landy-Szalay estimator \citep{LS},
\begin{equation}
w(\theta) = \frac{DD(\theta) - 2 DR(\theta) + RR(\theta)}{RR(\theta)}
\end{equation}
as implemented in the {\tt CUTE}
software\footnote{https://github.com/damonge/CUTE} \citep{2012arXiv1210.1833A}, where
$DD(\theta)$, $DR(\theta)$ and $RR(\theta)$ refer to normalized pair-counts of Data ($D$) and Random ($R$)
points, separated by an angular aperture $\theta$. Random points are uniformly
distributed across the footprint defined by our mask (albeit
downsampled following the fractional coverage of each pixel, described
in Sec.~\ref{sec:mask}), with an abundance twenty times larger than
that of the data in each given bin. For the fits and $\chi^2$ values
quoted in this section we always consider 16 angular-bins linearly
spaced between
$\theta=0.45\,{\rm deg}$ and $\theta=4.95\,{\rm deg}$, matching
the scale cuts in the BAO analysis using $w(\theta)$ of \main. 
We compute
pair-counts in angular aperture bins of width $0.3\,{\rm
  deg}$ in order to reduce the covariance between the
measurements.
The covariance matrix is derived from 1800 Halogen mocks, described in
detail in \mocks. 

The expected noise in the inverse covariance from the finite
number of realisations \citep{2007A&A...464..399H} and the translation of that into the variance of
derived parameters \citep{2013PhRvD..88f3537D} is negligible given the size of our data vector (16
angular measurements per tomographic redshift bin) and the number of
model parameters (one bias per bin). For instance the increased error
in derived best-fit biases in any given bin would be sub-percent.
The change in the full $\sqrt{\chi^2}$ is $\sim 3.7\%$ (16x4 data-points,
see the discussion below).
We therefore neglect these corrections in this section. 

\begin{figure}
\begin{center}
\includegraphics[width=0.9\linewidth]{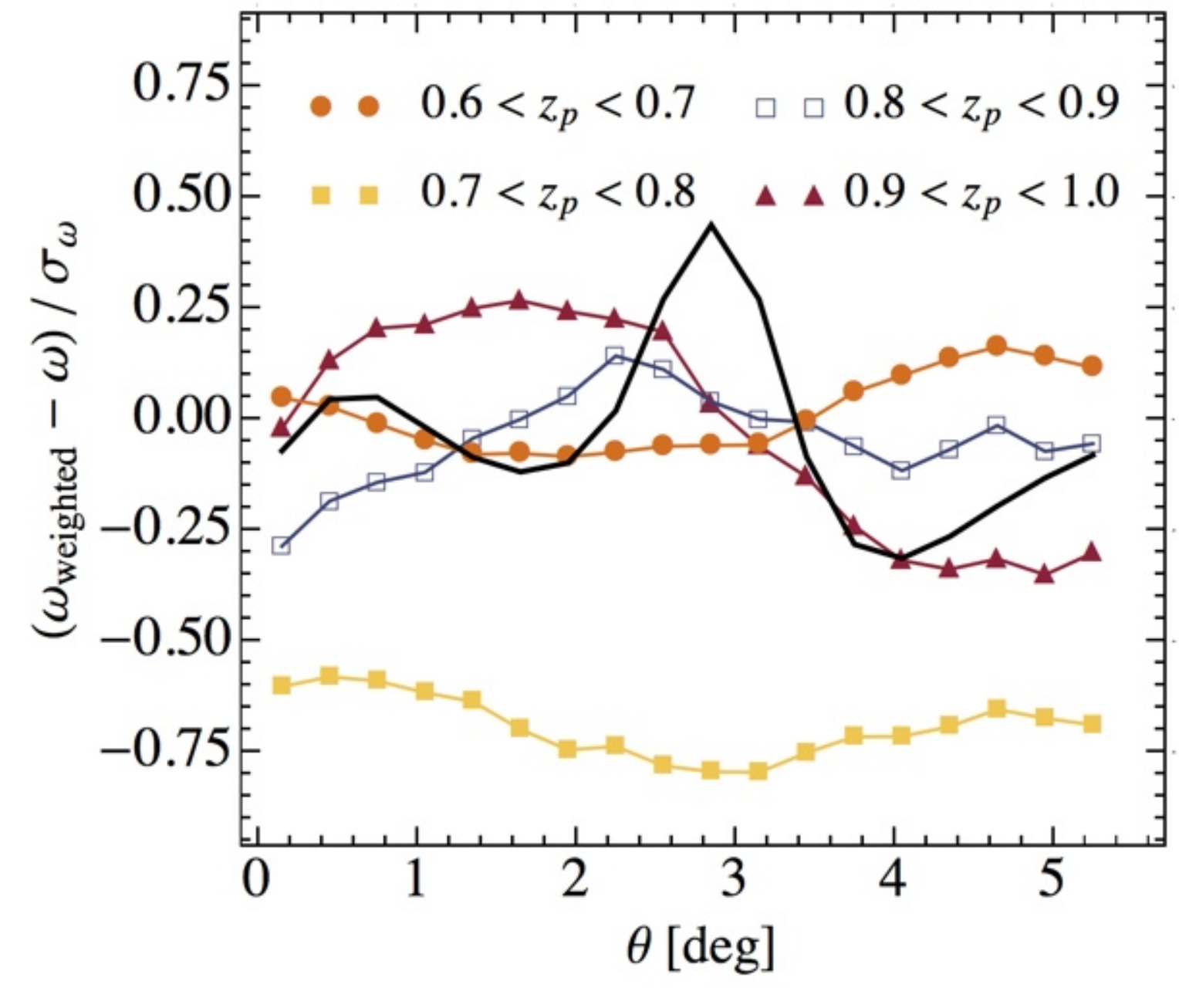} \\
\includegraphics[width=0.9\linewidth]{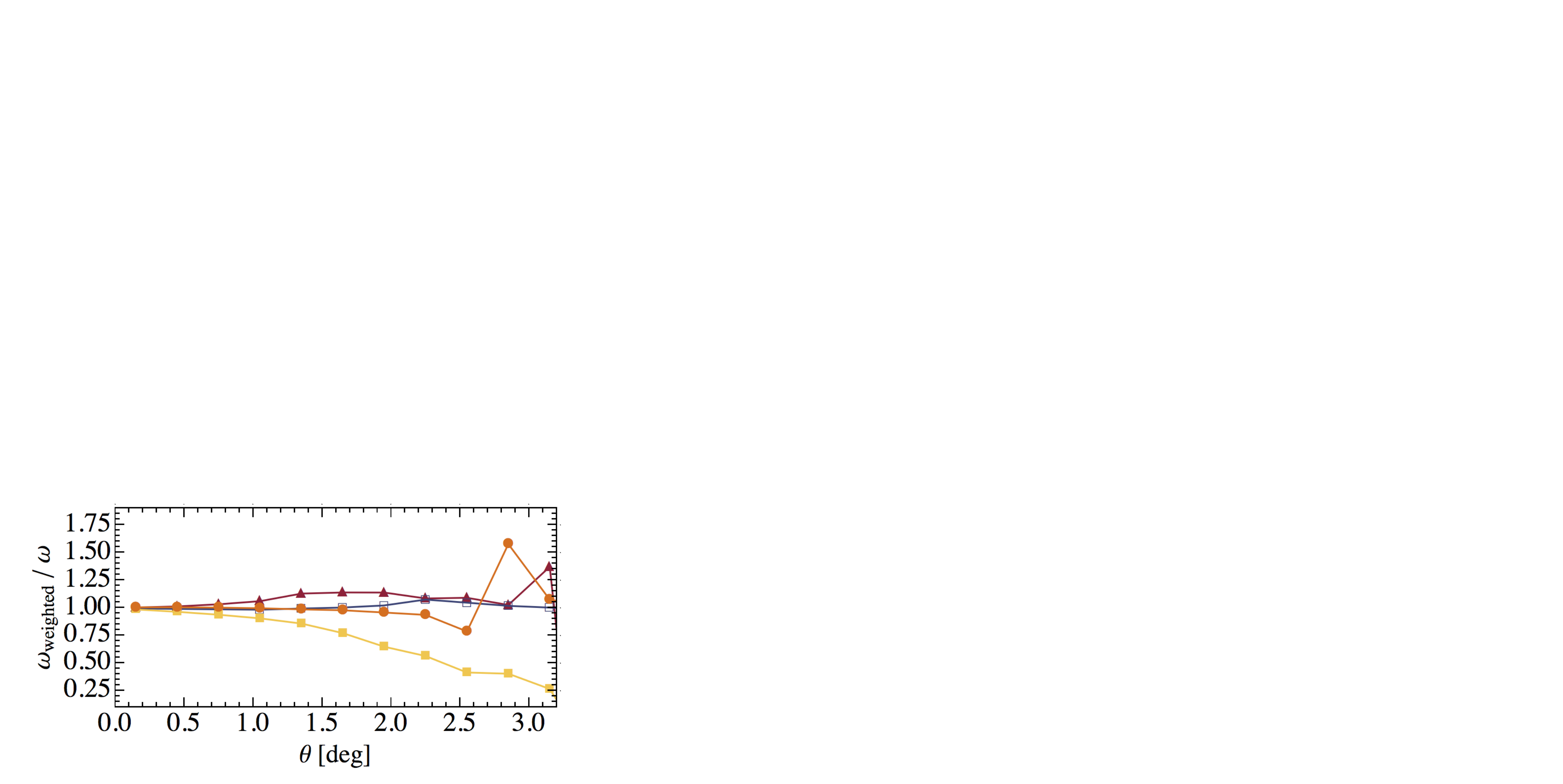}
\caption{Top panel shows the impact of the systematic weights on each redshift bin,
  shown by the differential angular correlations, with and without weights applied,
  relative to the uncertainty. One can see that the weights make the
  biggest difference for the $0.7 < z < 0.8$ bin, which is the
  redshift range with the greatest stellar contamination. The thick
  solid line displays the BAO feature in similar units, $( w_{\rm
    BAO}-w_{\rm no\,BAO} )/\sigma_w$, for the second tomographic bin as an
  example (different bins show similar BAO strength but displaced slightly
  in the angular coordinate). The systematic weights only modify the underlying smooth
  shape, and do not have a sharp feature at BAO scales. \change{Bottom panel
  shows the ratio of correlations for each bin, which provides
  additional information on the absolute size of the corrections (in this case we only plot up to
  scale with no zero crossings of $w$)}.} 
\label{Fig:wtheta_weights}
\end{center}
\end{figure}

\begin{figure*}
\includegraphics[width=1\linewidth]{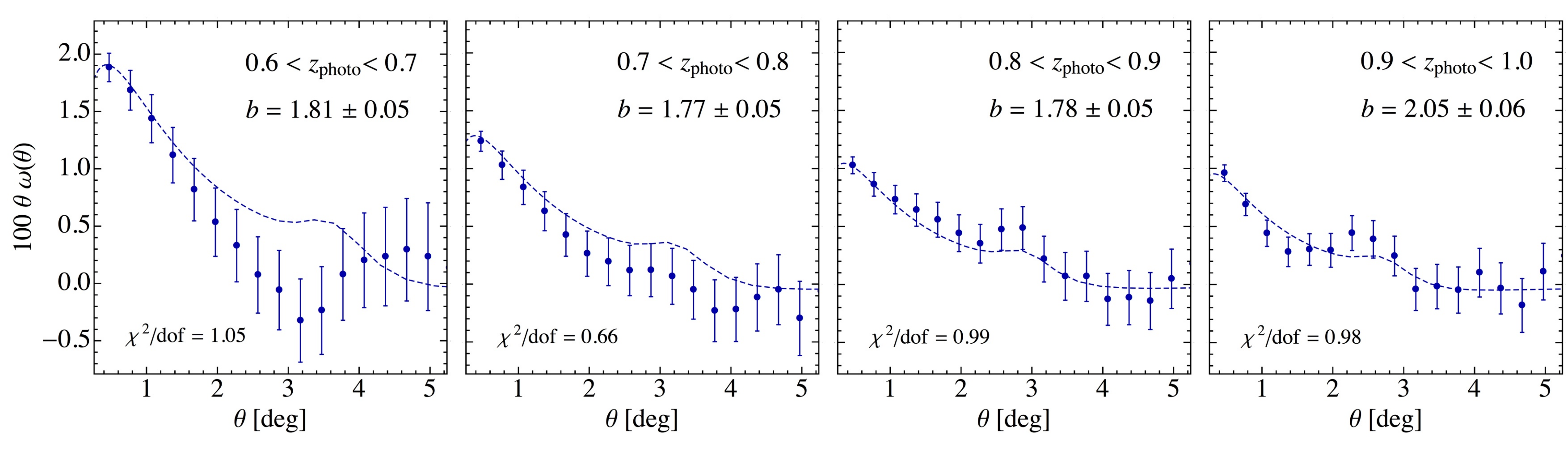}
\caption{Angular correlation function in four redshift bins, for
  galaxies selected with $z_{\rm DNF-MOF}$. Symbols
  with error bars show the clustering of galaxy sample
  corrected for the most relevant systematics. Dashed line displays a model using linear theory
  with an extra damping of the BAO feature due to nonlinearities,
  and a linear bias fitted to the data (whose best fit value is reported in the
  inset labels). We consider $16$
  data-points and one fitting parameter in each case
  (dof=$15$). Note that the points are very covariant, which might
  explain the visual mismatch in the first tomographic bin that
  nonetheless retains a good $\chi^2/{\rm dof}$.} 
\label{Fig:wtheta_clustering_auto}
\end{figure*}

\begin{figure*}
\includegraphics[width=0.75\linewidth]{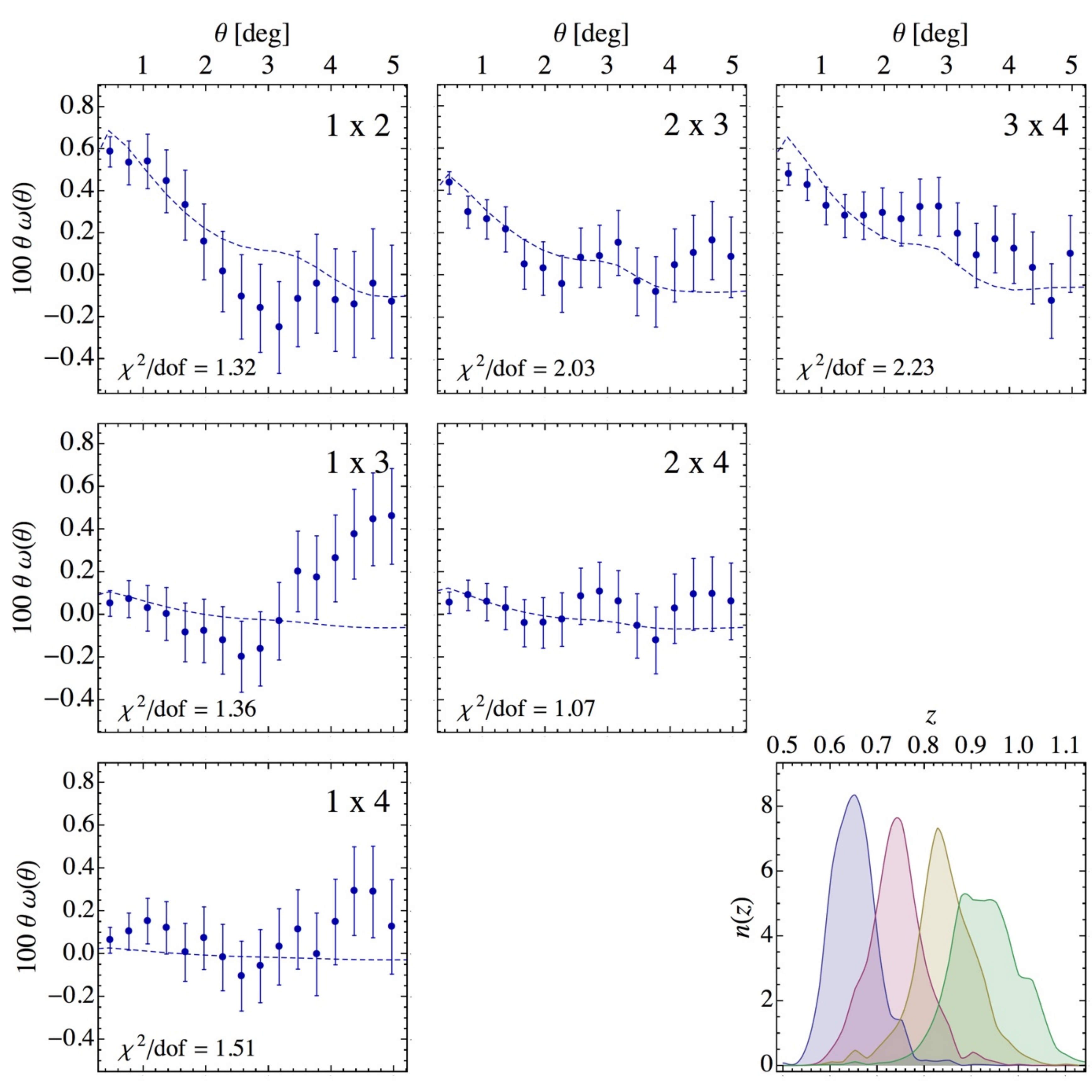}
\caption{Angular cross-correlation functions of the four
  tomographic bins in $0.6 < z_{\rm photo} < 1.0$, see Fig. \ref{Fig:wtheta_clustering_auto}, for galaxies selected
  according to $z_{\rm DNF-MOF}$. The model prediction shown with
  dashed lines
  assumes a bias equal to the geometric mean of the auto-correlation
  fits, i.e. $b_{ij}=\sqrt{b_i b_j}$, and is basically proportional to the
  overlap of redshift distributions, which are shown in the bottom right
  panel.} 
\label{Fig:wtheta_clustering_cross}
\end{figure*}

Figure \ref{Fig:wtheta_weights} shows the impact of the systematic
weights on the measured angular clustering in terms of the difference
$\Delta w$ between the pre-weighted correlation function $w$ and the
post-weighted one $w_{\rm weighted}$, relative to the statistical error
$\sigma_w$ (i.e. neglecting all covariance). To compare this against
the expected amplitude of the BAO feature at this scales we also display in
thick solid black line the theoretical angular correlation function
with and without BAO, for the second tomographic bin for
concreteness, relative to the
statistical errors. The corrections are all at the same level (or
smaller) than the expected BAO signal. 

The weights have the
largest impact in terms of clustering amplitude for the redshift bin $0.7 < z < 0.8$, which is the
redshift range with the largest stellar contamination ($\sim 4\%$, see
Table \ref{tab:sample}), although never exceeding one $\sigma_w$. For the remaining bins
the change in the correlation functions are within $1/4$ of $\sigma_w$.
We can assess quantitatively
the total potential impact of the weights 
 by calculating $\chi_{\rm sys}^2 =
\Delta w(\theta) ^{t} C^{-1} \Delta w(\theta)$; the square-root of this
number is an upper bound in the impact, in terms of number of $\sigma$'s, that the
weights could have on the determination of any model parameter.

 In the range $0.45\,{\rm deg} < \theta
< 4.95\,{\rm deg}$, with $16$ data-points, we find $\chi_{\rm sys}^2 = 0.1, 1.35, 0.2$
and $0.5$ respectively for each tomographic bin separately (showing
that for example best-fit bias derived solely from the 2nd tomographic
bin can be shifted by more than one sigma if weights are uncorrected for). More
interestingly, for the four bins combined and including the full
covariance matrix, we find $\chi_{\rm sys}^2 = 1.35$. This implies a maximum
impact of $1.16\sigma$ in a derived global parameter such as the angular
diameter distance measurement. This maximum threshold is well above the actual
impact of the weights in $D_A/r_s$ found in \main, which is $0.125 \sigma_{D_A/r_s}$ (see
Table 5 in that reference). We consider this an indication that the particular shape
of the BAO feature is not easily reproducible by contaminants, and is
therefore largely insensitive to such corrections, which is consistent
with previous analyses \citep{2017MNRAS.464.1168R}. 

Figure \ref{Fig:wtheta_clustering_auto} displays the auto-correlation function (including
observational systematic weights) of 4
tomographic bins of width $\Delta z_{photo} = 0.1$ between $0.6 \le z_{photo}
\le 1.0$.  Data at $z > 0.8$ appear to show significant BAO features.
Best fit biases, derived $1\sigma$ errors and their
corresponding $\chi^2$ values are reported as
inset panels and in Table \ref{tab:sample}. The model displayed
assumes linear theory and the MICE cosmology\footnote{We make this
  choice throughout the DES-Y1 BAO analysis because the MICE N-Body simulation was used to calibrate the
  Halogen mock galaxy catalogues. MICE cosmology assumes a 
  flat concordance LCDM model with $\Omega_{\rm matter}=0.25$, $\Omega_{\rm
    baryon}=0.044$, $n_s = 0.95$, $\sigma_8 = 0.8$ and $h=0.7$.}
\citep{2015MNRAS.448.2987F,2015MNRAS.453.1513C}, with an extra damping of the BAO feature, see
\methodACF\, for details. The $\chi^2/{\rm dof}$ are
all of order $\sim 1$ or better, showing that these are indeed good
fits given the covariance of the data. In Table \ref{tab:sample} we
also report best fit bias values for a split of the sample into four
tomographic bins using the ${\rm BPZ}_{\rm MOF}$ photo-$z$, showing no discrepancies.

As a further test of the clustering signal, as well as the
tails of the photo-$z$ distributions, we show in Fig. \ref{Fig:wtheta_clustering_cross} 
the cross-correlation between different bins. The overploted models
were derived using the redshift distributions of the corresponding
bins and assume a bias equal to the geometric mean of the tomographic
bins, 
\begin{equation}
\label{eq:wij}
w_{ij}(\theta) = b^2_{ij} \int \int \, dz \, d\tilde{z} n_i(z) n_j(\tilde{z}) D(z) D(\tilde{z}) \xi(r_\theta)
\end{equation}
where $r^2_\theta = r(z)^2 + r(\tilde{z})^2 - 2 r(z) r(\tilde{z}) \cos
\theta$ and $b^2_{ij} = b_i b_j$. In Eq.~(\ref{eq:wij}) we denote
$\xi$ the spatial correlation function computed in linear theory
at $z=0$. The error bar displayed and the reported $\chi^2$ values are
obtained with a theoretical covariance matrix designed to match the
Halogen mocks covariance of the auto-correlations (i.e. matching the
bias and shot noise and area of the mocks). Detailed formulae and
tests of this theory covariance are given in a companion paper,
\methodACF\, (see also \cite{2011MNRAS.414..329C,2011MNRAS.415.2193R,2014MNRAS.443.3612S}). However when we test the $\chi^2$ values of the
auto-correlations against the best-fit model\footnote{The best-fit
  bias and error from the theory covariance or the mocks one are
  consistent with each other, however the $\chi^2$ values are only so
  to about $40\%$.} using this theory covariance instead of the one
derived from the mocks
we find considerably larger $\chi^2$ values: $r_i \equiv \chi^2_{i, \rm theory-cov} /
\chi^2_{i, \rm mocks-cov} = 1.46, 1.37, 1.37, 1.47$ for auto-correlations
in bin $i=1$ to $4$, respectively. We propagate this uncertainty to the 
cross-correlations by dividing $\chi^2_{ij, {\rm theory-cov}}$ by $\sqrt{r_i r_j}$. 

Overall the cross-correlations show a good match to the model, which
is sensitive to the tails of the redshift distributions and the geometric
mean bias. The $\chi^2/{\rm dof}$ are $\sim 1$. The non-adjacent
bin $1 \times 3$ (where the expected clustering signal is negligible) shows
an excess correlation on very large-scales. This 
most probably indicate a residual systematic and not a problem of the
photo-z distributions. 

The large $\chi^2$ values in some of the cross-correlations (bins $2\times
3$ and $3\times 4$) are driven
by the non-diagonal structure of the covariance matrix rather than a
mismatch between the best-fit bias of the cross-correlation
$b_{ij}$ compared to the geometrical mean of the auto-correlation
biases. For example, for $2 \times 3$ the best-fit bias from
$w_{2\times 3}$ is only $2\%$ larger than $\sqrt{b_2 b_3}$ (and the corresponding $\chi^2$
change sub-percent). On the other hand, the $\chi^2$ of the
cross-correlation drops to $0.4$ if we only consider a diagonal
covariance matrix. Similarly $\chi^2_{3 \times 4}$ drops to 1.28 from $2$
  using a diagonal covariance matrix. Overall, we conclude there is a
  fairly good match between the implications of the overlap of
  redshift distributions and the cross-correlation clustering signal. 

\begin{figure}
\includegraphics[width=0.86\linewidth]{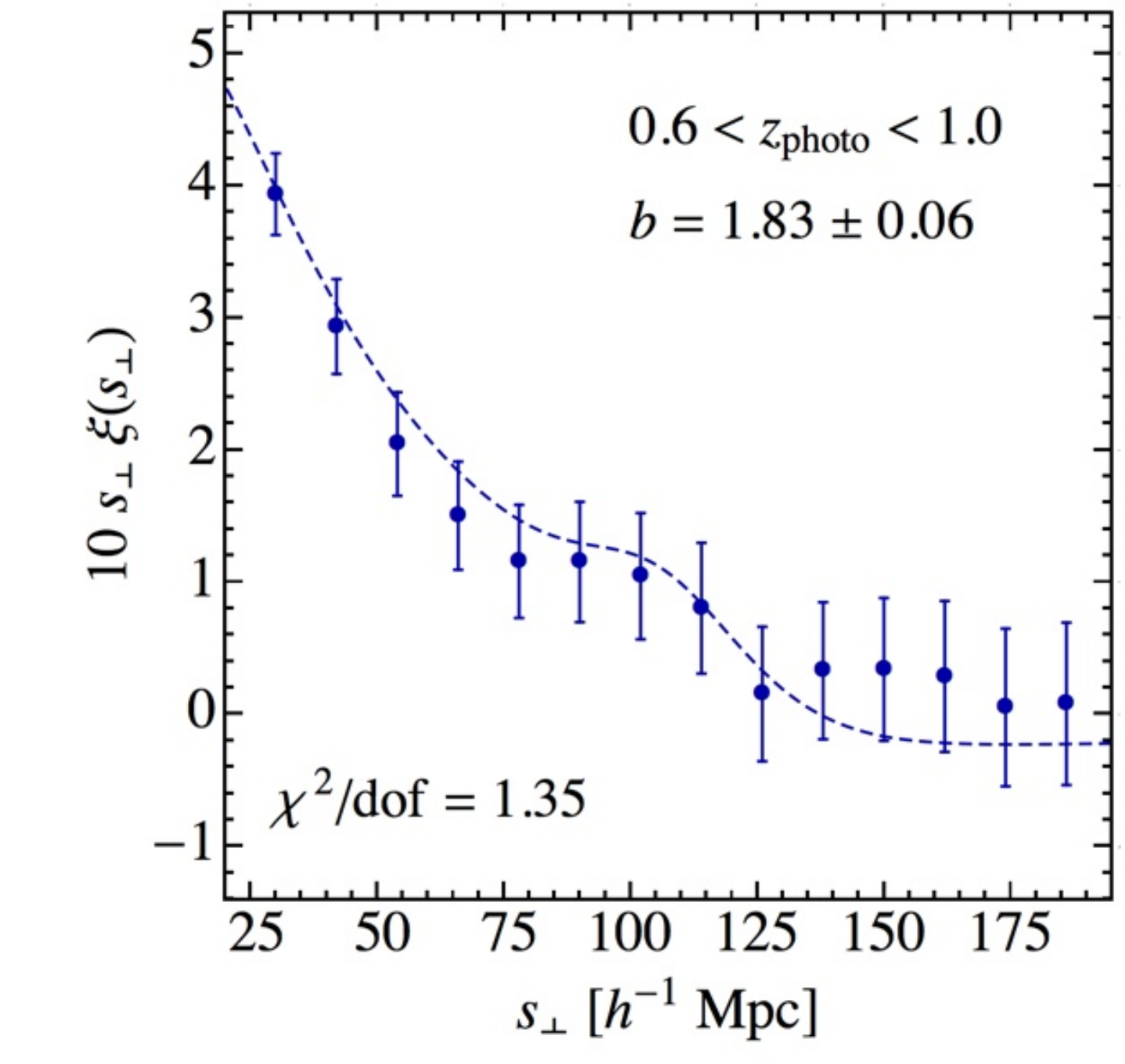}
\caption{Three-dimensional correlation function binned in projected
  pair separations. We use projected separations because radial pairs
  are damped due to photo-$z$ mixing. The dashed line is the
  best fit model assuming linear bias and a smeared BAO feature, as
  discussed in detail in \main.} 
\label{Fig:xi}
\end{figure}

In Figure \ref{Fig:xi} we show $\xi(s_{\rm perp})$ which is the three-dimensional correlation function binned only in projected physical
separations. To compute this correlation we converted (photometric) redshift and angles to
physical distances assuming MICE cosmology.
This yields a three-dimensional map of the galaxies in comoving coordinates. Random points
are distributed in this volume with the same angular distribution as
the angular mask defined in section \ref{sec:mask}, and used for $w(\theta)$, and drawing
redshifts randomly from the galaxies themselves. Pair counts are then
computed and binned in projected separations. A full detail of such procedure is
given in \main\, as well as in \cite{2017MNRAS.472.4456R}. The
modeling displayed in Fig.~\ref{Fig:xi} projects the real space three-dimensional correlation function into photometric space assuming Gaussian photometric
redshift errors per galaxy, provided in Table \ref{tab:sample} as $\sigma_{68}$. It also assumes a
linear bias betweeen the galaxies and the matter field.

The bias recovered from the three-dimensional projected clustering at
a mean redshift of $0.8$ is $b=1.83 \pm 0.06$, consistent with the one from $w(\theta)$ tomography.
In addition we
stress that this clustering estimate includes all 
cross-correlations of the data. The fact that it is matched by the theory modeling, which in turn
includes a characterisation of the redshift distributions per galaxy,
represents also an additional consistency check of reliability of the photometric redshifts.

\section{Conclusions} 
\label{sec:conclusions}

This paper describes the selection of a sample of galaxies, optimised for BAO 
distance measurements, from the first year of DES data. By construction,
this sample is dominated by red and luminous galaxies with redshifts in the 
range $0.6 < z < 1.0$. We have extended the selection of red galaxies beyond that
of previously published imaging data used for similar goals in SDSS by \cite{Pad05}
to cover the higher redshift and deeper data provided by DES.

We compute the expected magnitudes of galaxy templates in the four DES filters
and identify the ($i-z$) and ($z-i$) color space to select red galaxies in the
redshift range of interest. The actual selection in color and
magnitude is defined using the BAO distance measurement
figure-of-merit as a guiding criteria. Remarkably, the resulting
forecast matches the results obtained in \main\,with the final
analysis. The global flux limit of the sample is $i_{\rm auto} < 22$,
although we later introduce a sliding magnitude cut to limit ourselves
to brighter objects towards lower redshifts.

We consider three different photo-$z$ catalogues, with two different
photometric determinations. We showed that the typical photo-$z$ uncertainty (in units of $1+z$)
goes from $2.3\%$ to $3.6\%$ from low to high redshift, for DNF
redshifts using MOF photometry, and slightly worse for BPZ with MOF
photometry. Hence the former constitutes our primary catalogue in
\main, while the later is used for consistency. Redshift estimations
based on COADD photometry turned out to be worse than those derived
from MOF photometry by $10\%-20\%$.
Our final sample is made of 1.3 million red galaxies across 1336
deg$^2$ of area, largely contained in one compact region (SPT).

We study and mitigate, when needed, observational systematics traced
by various survey property maps. Of these, the most impactful is the
stellar contamination, which we find nonetheless bound to $<4\%$. 
Also $i$-band mean seeing and $g$-band depth are relevant.
We define weights to be applied to the galaxies when computing pair
counting to remove the relations between galaxy number density and large scale
fluctuations in those survey properties. 
We show that none of these corrections have an impact on BAO measurements, mainly
because they can eventually modify the broad-shape of the correlation
functions but do not introduce a characteristic localised scale as the
BAO. 

Lastly we characterised the two-point clustering of the sample, which
is then used in \main\,to derived distance constraints. We find the
auto-correlations to be consistent with a bias that evolves only
slightly with redshift, from $1.8$ to $2$. The bias derived from the
tomographic analysis is consistent with the one fitted to the whole
sample range with the 3D projected distance analysis. Furthermore we
investigate the cross-correlation between all the tomographic bins
finding clustering amplitudes matching expectactions, although with
poor $\chi^2$-values in some cases. Overall this is a
further test of the assumed redshift distributions. 

This paper serves the purpose of enabling for the fist time BAO distance measurements
using photometric data to redshifts $z \sim 1$. These measurements
achieve a precision comparable to those considered
state-of-the-art using photometric redshift to this point
\citep{2012ApJ...761...13S}, as well as those from WiggleZ \citep{2011MNRAS.415.2892B}, which are both limited to $z \sim
0.65$. These BAO
results are presetend in detail in \main. While this paper was completed, the
third year of DES data was made available to the collaboration,
totalling 3 to 4 times the area presented here, and similar or better depth.
Hence we look forward to that analysis, which should already yield a
very interesting counter-part to the high precision low-$z$ BAO
measurements already existing.

\section*{Acknowledgments}
MC acknowledges support from the Spanish Ramon y Cajal MICINN program. MC and EG have been partially funded by AYA2015-71825.
AJR is grateful for support from the Ohio State University Center for
Cosmology and AstroParticle Physics..
KCC acknowledges the support from the Spanish Ministerio de Economia y Competitividad grant ESP2013-48274-C3-1-P and the Juan de la Cierva fellowship. This work has made use of CosmoHub, see \cite{Carretero:2017zkw}. CosmoHub has been developed by the Port d'Informaci\'{o} Cient\'{i}fica (PIC), maintained through a collaboration of the Institut de F\'{i}sica d'Altes Energies (IFAE) and the Centro de Investigaciones Energ\'{e}ticas, Medioambientales y Tecnol\'{o}gicas (CIEMAT), and was partially funded by the ``Plan Estatal de Investigaci\'{o}n Científica y T\'{e}cnica y de Innovaci\'{o}n'' program of the Spanish government.

We are grateful for the extraordinary contributions of our CTIO colleagues and the DECam Construction, Commissioning and Science Verification
teams in achieving the excellent instrument and telescope conditions that have made this work possible.  The success of this project also 
relies critically on the expertise and dedication of the DES Data Management group.

Funding for the DES Projects has been provided by the U.S. Department of Energy, the U.S. National Science Foundation, the Ministry of Science and Education of Spain, 
the Science and Technology Facilities Council of the United Kingdom, the Higher Education Funding Council for England, the National Center for Supercomputing 
Applications at the University of Illinois at Urbana-Champaign, the Kavli Institute of Cosmological Physics at the University of Chicago, 
the Center for Cosmology and Astro-Particle Physics at the Ohio State University,
the Mitchell Institute for Fundamental Physics and Astronomy at Texas A\&M University, Financiadora de Estudos e Projetos, 
Funda{\c c}{\~a}o Carlos Chagas Filho de Amparo {\`a} Pesquisa do Estado do Rio de Janeiro, Conselho Nacional de Desenvolvimento Cient{\'i}fico e Tecnol{\'o}gico and 
the Minist{\'e}rio da Ci{\^e}ncia, Tecnologia e Inova{\c c}{\~a}o, the Deutsche Forschungsgemeinschaft and the Collaborating Institutions in the Dark Energy Survey. 

The Collaborating Institutions are Argonne National Laboratory, the University of California at Santa Cruz, the University of Cambridge, Centro de Investigaciones Energ{\'e}ticas, 
Medioambientales y Tecnol{\'o}gicas-Madrid, the University of Chicago, University College London, the DES-Brazil Consortium, the University of Edinburgh, 
the Eidgen{\"o}ssische Technische Hochschule (ETH) Z{\"u}rich, 
Fermi National Accelerator Laboratory, the University of Illinois at Urbana-Champaign, the Institut de Ci{\`e}ncies de l'Espai (IEEC/CSIC), 
the Institut de F{\'i}sica d'Altes Energies, Lawrence Berkeley National Laboratory, the Ludwig-Maximilians Universit{\"a}t M{\"u}nchen and the associated Excellence Cluster Universe, 
the University of Michigan, the National Optical Astronomy Observatory, the University of Nottingham, The Ohio State University, the University of Pennsylvania, the University of Portsmouth, 
SLAC National Accelerator Laboratory, Stanford University, the University of Sussex, Texas A\&M University, and the OzDES Membership Consortium.

Based in part on observations at Cerro Tololo Inter-American Observatory, National Optical Astronomy Observatory, which is operated by the Association of 
Universities for Research in Astronomy (AURA) under a cooperative agreement with the National Science Foundation.

The DES data management system is supported by the National Science Foundation under Grant Numbers AST-1138766 and AST-1536171.
The DES participants from Spanish institutions are partially supported by MINECO under grants AYA2015-71825, ESP2015-66861, FPA2015-68048, SEV-2016-0588, SEV-2016-0597, and MDM-2015-0509, 
some of which include ERDF funds from the European Union. IFAE is partially funded by the CERCA program of the Generalitat de Catalunya.
Research leading to these results has received funding from the European Research
Council under the European Union's Seventh Framework Program (FP7/2007-2013) including ERC grant agreements 240672, 291329, and 306478.
We  acknowledge support from the Australian Research Council Centre of Excellence for All-sky Astrophysics (CAASTRO), through project number CE110001020.

This manuscript has been authored by Fermi Research Alliance, LLC under Contract No. DE-AC02-07CH11359 with the U.S. Department of Energy, Office of Science, Office of High Energy Physics. The United States Government retains and the publisher, by accepting the article for publication, acknowledges that the United States Government retains a non-exclusive, paid-up, irrevocable, world-wide license to publish or reproduce the published form of this manuscript, or allow others to do so, for United States Government purposes.

This paper has gone through internal review by the DES collaboration.
The DES publication number for this article is DES-2017-0305. The Fermilab pre-print number is FERMILAB-PUB-17-585.

\bibliographystyle{mn2e}
\bibliography{biblio}
\renewcommand{\thesubsection}{\Alph{subsection}}

\label{lastpage}

\end{document}